\documentclass[prd,aps,showpacs,twocolumn,superscriptaddress,floatfix]{revtex4}

\usepackage{newtxtext,newtxmath}
\usepackage[T1]{fontenc}
\usepackage{ae,aecompl}
\usepackage{dcolumn}
\usepackage{bm}
\usepackage{amsmath}
\usepackage{float}
\usepackage[utf8]{inputenc}
\usepackage{pgfplots}
\usepackage{xcolor}
\pgfplotsset{compat=1.13}
\usepackage{graphicx}
\usepackage{natbib}
\usepackage{aas_macros}
\usepackage{hhline}

\usepgfplotslibrary{groupplots}

\definecolor{capri}{rgb}{0.0, 0.75, 1.0}

\begin{document}

\title{Empirical relations for gravitational-wave asteroseismology of binary neutron star mergers}


\author{Stamatis Vretinaris}
\email{svretina@physics.auth.gr}
 \affiliation{Department of Physics, Aristotle University of Thessaloniki, University Campus, 54124, Thessaloniki, Greece.}
\author{Nikolaos Stergioulas}%
 \email{niksterg@auth.gr}
\affiliation{Department of Physics, Aristotle University of Thessaloniki, University Campus, 54124, Thessaloniki, Greece.}%

\author{Andreas Bauswein}
\email{a.bauswein@gsi.de}
\affiliation{GSI Helmholtzzentrum f{\"u}r Schwerionenforschung, Planckstra{\ss}e 1, 64291 Darmstadt, Germany.}
\affiliation{Heidelberg Institute for Theoretical Studies, Schloss-Wolfsbrunnenweg 35, 69118 Heidelberg, Germany.}%

\date{\today}

\begin{abstract}
We construct new, multivariate empirical relations for measuring neutron star radii and tidal deformabilities from the dominant gravitational wave frequency in the post-merger phase of binary neutron star mergers. The relations determine neutron star radii  and tidal deformabilities for specific neutron star masses with consistent accuracy and depend only on two observables: the post-merger peak frequency $f_{\rm peak}$ and the chirp mass $M_{\rm chirp}$. The former could be measured with good accuracy from gravitational waves emitted in the post-merger phase using next-generation detectors, whereas the latter is already obtained with good accuracy from the inspiral phase with present-day detectors. Our main data set consists of a gravitational wave catalogue obtained with CFC/SPH simulations. We also extract the $f_{\rm peak}$ frequency from the publicly available CoRe data set, obtained through grid-based GRHD simulations and find good agreement between the extracted frequencies of the two data sets. As a result, we can construct empirical relations for the combined data sets. Furthermore, we investigate empirical relations for two secondary peaks, $f_{2-0}$ and $f_{\rm spiral}$, and show that these relations are distinct in the whole parameter space, in agreement with a previously introduced spectral classification scheme.
Finally, we show that the spectral classification scheme can be reproduced using machine-learning techniques.\end{abstract}

\maketitle

\section{Introduction}


The first detection of gravitational waves (GWs) from a binary neutron star (BNS) merger, GW170817, in August of 2017 signaled the beginning of the era of GW-multimessenger astronomy \cite{Abbott2017}. Finite-size effects during the pre-merger phase lead to constraints on the tidal deformability~\cite{Hinderer2010} (which is related to the radius) directly from the GW signal~\cite{Abbott2017,Chatziioannou2018,PhysRevLett.121.161101,De2018,Carney2018}. With additional detections expected in the forthcoming observation runs of the aLIGO/aVIRGO detectors it is expected that constraints from the inspiral phase will gradually tighten e.g.\ \cite{Read2013,DelPozzo2013,Wade2014,Agathos2015,Chatziioannou2015,Hotokezaka2016,Chatziioannou2018}. 

Employing a multi-messenger interpretation of GW170817, i.e.\ exploiting additional information from the electromagnetic counterpart, additional constraints on neutron-star (NS) parameters were derived including a robust lower bound on NS radii~\cite{Margalit2017,Bauswein2017,Shibata2017,Rezzolla2018,Radice2018,Ruiz2018,Coughlin2018,Koeppel2019,Kiuchi2019,Capano2019,2019AIPC.2127b0013B,BAUSWEIN2019167958}. Constraints on NS radii and the tidal deformability can be directly translated to constraints on the high-density part of the NS equation of state (EOS) \cite{Fattoyev2017,Raithel2018,PhysRevLett.121.161101,2019arXiv190511212T,2019arXiv190605978C}.

Another method for directly measuring NS radii is through observations of the {\it postmerger} phase of BNS mergers (see~\cite{Bauswein2012,Bauswein2012a} for initial publications and \cite{2019arXiv190106969B,2019arXiv190708534B} for recent extensive reviews and references therein). For GW170817 the GW instruments were not yet sufficiently sensitive to detect GW emission from the postmerger phase~\cite{Abbott2017a}, but measurements can be anticipated when the detectors reach design sensitivity or when projected upgrades are installed, e.g.~\cite{Torres-Rivas2019}. For typical NS masses, this method is complementary to measuring the tidal deformability in the inspiral phase,  but it also has the potential of placing even tighter constraints on the radius of massive NS, the maximum mass of nonrotating NSs, the tidal deformability or to probe the existence of a quark core~\cite{Bauswein2012a,Bauswein2013,Bauswein2014a,CORE1,Most2019,Bauswein2019a}. This is because the remnant in the postmerger phase reaches higher maximum densities that are inaccessible by methods which consider the relatively light progenitor stars before merging.

The remnant of a BNS merger that has a sufficiently low mass to avoid prompt collapse is a stable or meta-stable differentially rotating NS remnant, whose dynamics are influenced mainly by the EOS, the total binary mass and the mass ratio. Gravitational waves emitted in the post-merger phase contain quasi-discrete, long-lived frequency components, as well as short-lived initial transients, e.g.~\cite{Shibata2005a,Stergioulas2011,Bauswein2015,Takami2015,Paschalidis2015,Clark2016,Foucart2016,Rezzolla2016,Radice2016a,Maione2017}. These originate from specific mechanisms that are sensitive to the EOS. By relating the post-merger spectrum to properties of individual NSs one can constrain the EOS.

Specifically,  the postmerger spectrum has several distinct peaks in the kHz regime which are produced by certain physical mechanisms connected to oscillation modes and dynamical features of the postmerger remnant. The dominant oscillation frequency $f_{\mathrm{peak}}$ in the GW spectrum is a generic feature, which occurs in all merger simulations that do not result in a prompt collapse \cite{Shibata2005}. The underlying mechanism that produces this frequency is the excitation of the \textit{fundamental quadrupolar fluid mode l=m=2}, as shown in~\cite{Stergioulas2011,Bauswein2016}. 

At frequencies somewhat smaller than $f_{\mathrm{peak}} $ two additional, potentially detectable secondary peaks can appear, $f_{\mathrm{2-0}}$ and $f_{\mathrm{spiral}}$~\cite{Shibata2005a,Stergioulas2011,Bauswein2015,Bauswein2016}. If we denote the frequency of the fundamental quasi-radial mode of the remnant as $f_0$ (which itself produces extremely weak GW emission), then  \textit{quasi-linear combination frequencies} $f_{2\pm 0} = f_2 \pm f_0$ are present in the GW spectrum (where $f_2 \equiv f_{\mathrm{peak}}$). The existence of such combination frequencies is a natural consequence of the nonlinear nature of the evolution of simultaneous oscillations in the remnant. In some models, the $f_{2-0}$ peak is potentially detectable, while in others it is suppressed, due to a strong damping of the postmerger quasi-radial oscillations~\cite{Bauswein2015}.

The other secondary peak, $f_{\rm spiral}$  occurs at frequencies between $f_{\mathrm{2-0}}$ and $f_{\mathrm{peak}}$~\cite{Bauswein2015}. This secondary peak is generated by the orbital motion of two antipodal bulges that form at the surface of the remnant after the merging, due to a tidal deformation, which has a spiral form in the case of equal-mass remnants.  Matter in the two antipodal bulges orbits around the remnant with an orbital frequency smaller than the pattern speed of the $l=m=2$ $f-$mode oscillating in the inner region.  This is a transient feature that lasts only for a few milliseconds.

Bivariate empirical relations between the dominant postmerger frequency $f_\mathrm{peak}$ and EOS properties were first investigated for fixed binary mass configurations varying the total mass and the mass ratio \cite{Bauswein2012,Bauswein2012a}. Stellar parameters of nonrotating NSs are uniquely linked to the EOS through the Tolman-Oppenheimer-Volkoff (TOV) equations. For example, the peak frequency $f_\mathrm{peak}$ of 1.35-1.35~$M_\odot$ mergers shows a clear correlation with the radius $R_{1.35}$ of a nonrotating NS with 1.35~$M_\odot$ (see Fig.~4 in \cite{Bauswein2012} and Fig.~12 in \cite{Bauswein2012a}). Similar tight correlations exist for other fiducial masses  (see Figs.~9 to 12 in \cite{Bauswein2012a}). The tightest relation for a 1.35-1.35~$M_\odot$ is with the radius $R_{1.6}$. This relation can be written as
\begin{equation}
f _ { \mathrm { peak } } = \left\{ \begin{array} { l l } 
           { - 0.2823 \cdot R _ { 1.6 } + 6.284, } & { \textrm { for } f _ { \rm peak  } < 2.8 \mathrm { kHz } }, \\ 
           { - 0.4667 \cdot R _ { 1.6 } + 8.713, } & { \textrm { for } f _ { \rm peak  } > 2.8 \mathrm { kHz } }, \end{array} \right.
\end{equation}
(the maximum deviation of the data points from a least-square fit is considered as figure of merit to assess the quality and accuracy of the relations). For $R_{1.6}$ the maximum scatter is less than 200~m. 

For other fixed binary masses, e.g.\ 1.2-1.2~$M_\odot$, 1.2-1.5~$M_\odot$ or 1.5-1.5~$M_\odot$ mergers, similar scalings between $f_\mathrm{peak}$ and NS radii exist~\cite{Bauswein2012a} and a single relation, scaled by the total mass is \cite{Bauswein2016}
\begin{equation}
f _ { \rm peak  } / M _ { \mathrm { tot } } = 0.0157 \cdot R _ { 1.6 } ^ { 2 } - 0.5495 \cdot R _ { 1.6 } + 5.5030,
\end{equation}
(see~\cite{CORE1} for a similar rescaling but with the tidal coupling constant).

For nonrotating stars it is known that the fundamental quadrupolar oscillation mode roughly scales with the mean density $\sqrt{M/R^3}$ \cite{Andersson1998}. For fixed-mass sequences a strong radius dependence may thus be expected. Since the mass of merger remnants typically exceeds the maximum mass of nonrotating NSs, the oscillation frequencies of the remnant cannot be directly connected to oscillation modes of a nonrotating NS of the same mass. But, the corrections by rotation and the extrapolation to higher masses are likely to depend in a continuous manner on the EoS. A detailed investigation of oscillation modes of differentially rotating merger remnants is still to be developed, but quasi-normal modes for uniformly rotating stars in full general relativity have already been calculated \cite{2019arXiv191008370K}.  A tentative explanation of the relations between $f_\mathrm{peak}$ and NS properties is presented in \cite{Chakravarti2019}. For a detailed summary of the work leading to the present publication see the review article \cite{2019arXiv190106969B}.

Here, we extend the scaled bivariate empirical frequency-radius relations of \cite{Bauswein2016} to multivariate relations, by including the dependency on the chirp mass $M_{\rm chirp}$ of binary systems. This dependency of frequency on both the radius and chirp mass is expanded to second order, yielding accurate empirical relations over a wide range of masses. The procedure is repeated for the secondary peaks, demonstrating a clear distinction between $f_{2-0}$ and $f_{\rm spiral}$, in agreement with \cite{Bauswein2015}. For $f_{\rm peak}$, we also construct the inverse multivariate empirical relations, which describe the radius as a function of $f_{\rm peak}$ and  $M_{\rm chirp}$, again with terms expanded up to second order. These inverse relations can be implemented directly in the data analysis of GW searches~\cite{Clark2014,Clark2016,Chatziioannou2017,Bose2018,Yang2018,Torres-Rivas2019} and show a consistency in determining the radius over a wide range of neutron star masses. 

Moreover, we employ a machine-learning algorithm to corroborate the existence of distinct classes of postmerger spectra, depending on the strength and presence of the different secondary GW features. The algorithm  detects three different types of postmerger spectra, fully in line with the spectral classification scheme introduced in~\cite{Bauswein2015}, where  three classes of postmerger GW spectra were manually identified depending on presence or absence of $f_{2-0}$ and $f_{\rm spiral}$.

Constraints on the high-density EOS can also be set by inferring the tidal deformability of neutron stars in the inspiral phase (see ~\cite{Abbott2017,Chatziioannou2018,TheLIGOScientificCollaboration2018a,PhysRevLett.121.161101,De2018,Carney2018} as well as \cite{2019arXiv190708534B} and references therein). 
On the other hand, when using gravitational waves generated by the postmerger
oscillations, the EOS constraints obtained through the inference of the tidal deformability 
should be nearly (but not exactly) equivalent to EOS constraints obtained through the inference of radii. In addition to our empirical relations for radii, we thus construct multivariate empirical relations also for tidal deformabilities.

In \cite{CORE1}, a bivariate empirical
relation between  $f_{\rm peak} M_{\rm tot}$ and the dimensionless quadrupole
tidal coupling constant $\kappa_{2}^{\mathrm{T}}$ was found (see also \cite{Takami2015,Rezzolla2016}), whereas in \cite{2019PhRvD.100d4047T} a similar relation in terms the mass-weighted tidal deformability
 $\tilde \Lambda$ (adjusted for the mass dependence) was constructed.
The multivariate relations we construct are of the form $\Lambda_{\rm x}(M_{\rm chirp}, f_{\rm peak})$, where $ \Lambda_{\rm x}$ is the dimensionless tidal deformability at a specific mass (indicated as the subscript $'x'$ in solar masses). These relations
depend only on
quantities that are directly measurable from the gravitational wave signal and are of significantly better accuracy than corresponding bivariate relations. 

The paper is organized as follows. In Sect.~\ref{sec:data} we summarize the data which we use for constructing empirical relations. Then we describe fits for postmerger GW frequencies in Sect.~\ref{sec:freq}. Sect.~\ref{sec:machine} discusses the results of machine-learning algorithms, which we employ for the identification of different types of postmerger spectra. In Sect.~\ref{sec:rad} empirical relations for the determination of NS radii from measured postmerger frequencies are presented. We describe an application of these relations for constraining the mass-radius relation of neutron stars in Sect.~\ref{sec:mr}. The validation of the empirical relations using an independent data set is discussed in Sect.~\ref{CORE:relations}. Sec. \ref{sec:fpeakL} presents empirical relations for the dominant postmerger frequency in terms of tidal deformabilities and Sect. \ref{sec:Lfpeak} discusses the inverse relations. We close with a discussion and our conclusions in Sect.~\ref{sec:sum}.

Throughout the text, \textit{all frequencies} in empirical relations and figures are given in units of kHz and \textit{all masses} are given in units of $M_\odot$ and refer to the gravitational mass (for binary systems at infinite orbital separation). Radii refer to the circumferential radius.
\begin{figure*}[ht]
\includegraphics[width=17cm]{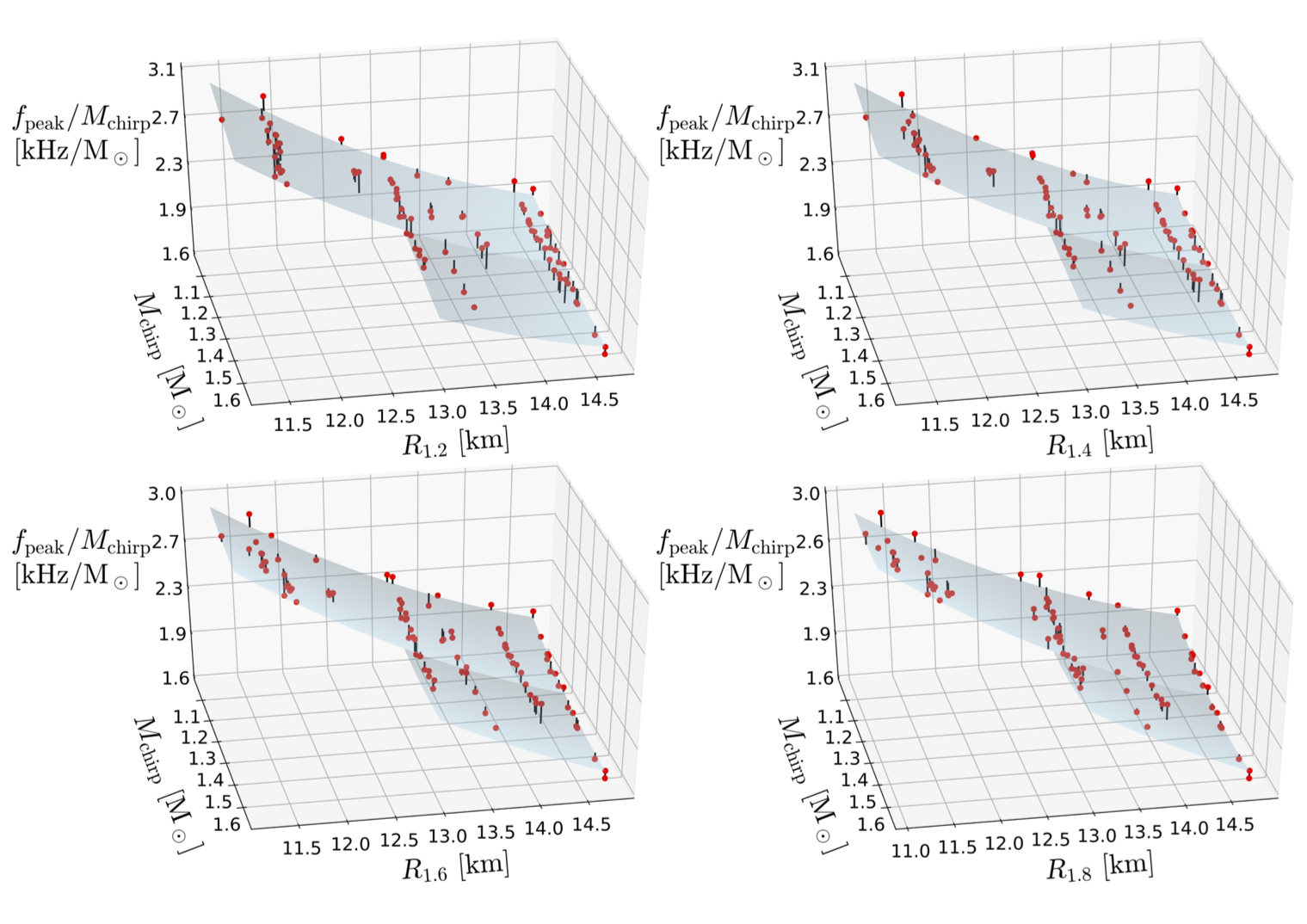}
  \caption{Surfaces $f_{\mathrm{peak}}(R_{\mathrm{x}},M_{\mathrm{chirp}})$ using the whole SPH/CFC data set. Red dots show the extracted frequencies $f_{\mathrm{peak}}$ scaled by the chirp mass $M_{\rm chirp}$ (in units of kHz/$M_\odot$), while the light blue surface represents the empirical relations of the form of Eq. (\ref{fRM}).  In the different panels, the radius of nonrotating neutron stars of mass 1.2, 1.4, 1.6 and 1.8$M_\odot$ was used. The surfaces are shown only in regions where data points are available.}
\label{fRMsurfaces}
\end{figure*}

\section{Data sets}\label{sec:data}

We construct empirical relations for the main post-merger GW frequencies using two different catalogues of GW waveforms. We start with 90 waveforms produced by a smoothed-particle (SPH) hydrodynamics code~\cite{Bauswein2012a,Bauswein2013a,Bauswein2014a,Bauswein2015} in the general-relativistc spatial conformal flatness (CFC) approximation~\cite{Wilson1996,Isenberg1980}. After establishing the new empirical relations, we use 28 waveforms of the publicly released CoRe data set~\cite{CORE} (which were produced by simulations in full general relativity and with high-resolution shock-capturing methods) to confirm the validity and accuracy of the new empirical relations. Finally, we produce empirical relations based on the combined data sets. 

\subsection{CFC/SPH GW catalogue}
\label{CFC-SPH}
Our first GW catalogue of BNS mergers is produced with a 3D SPH code~\cite{Oechslin2002,Oechslin2007,Bauswein2010a,Bauswein2012a}, which employs the CFC approximation for the evolution of the spacetime~\cite{Wilson1996,Isenberg1980}. Gravitational waves are extracted through a modified version of the quadrupole formula~\cite{Oechslin2007}. Both temperature-dependent EOSs and cold, barotropic models with an approximate treatment of thermal effects (see~\cite{Bauswein2010} for details and an assessment of this approximation in the context of BNS) are used. There are 49 equal mass models, with masses ranging from 1.2 $M_\odot$ to 1.9 $M_\odot$ and 41 unequal mass models, with masses ranging from 1.2 $M_\odot$ to 2.0 $M_\odot$ and mass ratios as low as 0.67. A summary of the main  properties of this catalogue is given in Appendix \ref{Appendix.D} in Table \ref{table:cfc/sph-data} and in Appendix \ref{Appendix.A} in Figs.  \ref{fig:eoschirp} and \ref{fig:eosconf}.

Before Fourier transforming the time domain data, we applied a Tukey window with a rolloff parameter $\alpha = 0.1$ and zero padded each time series to 16384 samples in total. We construct the effective amplitude $h_{eff} = \tilde{h} \sqrt{f}$, where $\tilde{h}$ is the Fourier transform of the time domain GW signal, from which individual frequency peaks are extracted. 

The extraction of the dominant postmerger frequency $f_{\rm peak}$ is always unambiguous, since it is the peak with the highest effective amplitude in the postmerger phase. For the extraction and identification of the two secondary peaks $f_{2-0}$ and $f_{\rm spiral}$ we use the spectral classification scheme introduced in \cite{Bauswein2015}, which distinguishes three different types of postmerger spectra: Type I , where $f_{2-0}$ dominates over $f_{\rm spiral}$, Type II, where $f_{2-0}$ and $f_{\rm spiral}$ are roughly comparable in amplitude, and Type III, where $f_{\rm spiral}$ dominates over $f_{2-0}$. The occurrence of the different types depends in a systematic way on the EOS and the binary masses. Specifically, the $f_{2-0}$ frequency can be found in the range $f_{\rm peak}-1.3$kHz to  $f_{\rm peak}-0.9$kHz (except for models very near the threshold mass to collapse, where the quasi-radial frequency diminishes) whereas the $f_{\rm spiral}$  frequency can be found in the range  $f_{\rm peak}-0.9$kHz to  $f_{\rm peak}-0.5$kHz. In the cases where a model is of Type I ($f_{2-0}$ dominates over $f_{\rm spiral}$) or Type III ($f_{\rm spiral}$ dominates over $f_{2-0}$) the correct identification of the main secondary frequency is straightforward. In a small number of mainly Type IIs cases, where $f_{2-0}$ and $f_{\rm spiral}$ are of comparable amplitude, some further considerations were required (for example, extraction of the quasi-radial frequency from the hydrodynamical simulation) in order to correctly identify the secondary peaks.

In order to relate the postmerger GW frequencies to the radius of individual nonrotating stars, we computed nonrotating models of different masses with the same set of EOSs as for the BNS merger simulations. For EOSs that are defined as a piecewise polytropes in \cite{Read2009}, we used the {pyTOVpp code} \footnote{The python code {\tt pyTOVpp} was used, available at \protect\url{https://github.com/niksterg/pyTOVpp} .}, whereas other EOSs were implemented in their original tabulated form with the RNS code \cite{Stergioulas1995}. 
Small discrepancies that arise in the determination of the radius of a nonrotating star between the tabulated and the piecewise polytropic approximation of an EOS are within the maximum deviation of the empirical relations.

\subsection{CoRe GW catalogue}

The CoRe GW catalogue \cite{CORE} is a large public database of BNS waveforms constructed through simulations in  full numerical relativity. We selected a subset of models for which the initial stars have zero spin and eccentricity lower than 0.02. In cases where the same model is available for multiple resolutions, we selected the highest resolution (denoted as $R01$ in \footnote{\protect\url{http://www.computational-relativity.org}}). Also, in cases where multiple waveforms were available for initial setups that differed only slightly in mass (due to a different initial separation distance), we selected the model with the lowest initial GW frequency (at the start of the simulation, before merger) which corresponds to the largest initial seperation distance. The subset of models we \textit{selected} from the CoRe GW catalogue are described in more detail in Appendix \ref{Appendix.B}, in Figs. \ref{fig:COREeoschirp} and \ref{fig:COREmodels} and in Appendix \ref{Appendix.D} in Table \ref{table:CORE-data}. This subset includes equal-mass models in the mass range 1.35 $M_\odot$ to 1.5 $M_\odot$ and unequal mass models in the mass range 0.94 $M_\odot$ to 1.94 $M_\odot$ and with a mass ratio as low as 0.49. There are 6 different EOS in this subset of selected models (compared to 13 different EOS in the Bauswein et al. catalogue described in Sec. \ref{CFC-SPH}). It also covers a smaller range of chirp masses, $1.06-1.2$, compared to $1.04-1.65$ in the CFC/SPH GW catalogue, but a larger range of mass ratios, $0.49-1.0$, compared to $0.67-1.0$ in the CFC/SPH GW catalogue. For more detailed information on the specific models we selected see the references \cite{CORE1,CORE2,CORE3,CORE4,CORE5,CORE6,CORE7,CORE8}. 

For our seleced subset of models from the  CoRe catalogue we only extracted the dominant $f_{\mathrm{peak}}$ frequency, in the same way as described in Sec. \ref{CFC-SPH}. These frequencies are then used in Sec. \ref{CORE:relations} to validate the empirical relations constructed with the CFC/SPH  GW catalogue, but also to construct empirical relations for the combined data set (i.e.\  combination of the Bauswein et al. data and the selected subset of the CoRe catalogue) in Sec. \ref{CORE:relations}.

\section{Empirical relations FOR\ FREQUENCIES\ based on the CFC/SPH catalogue}\label{sec:freq}
Using a least-squares minimization method \footnote{The python package {\tt Lmfit} was used, avalable at \protect\url{https://lmfit.github.io/lmfit-py/} . } (see \cite{lmfit}), we construct two-parameter relations of the form $f_j(R_{\rm x}, M_{\rm chirp})$, where $j$ stands for one of the three frequency peaks $f_{\rm peak}$, $f_{2-0}$ or $f_{\rm spiral}$ and ${\rm x}$ stands for the mass of fiducial nonrotating NS models, in solar masses (e.g.\ $R_{1.6}$ stands for the radius of a nonrotating model of mass $M=1.6 M_\odot$).

$M_{\rm chirp}$ is the usual chirp mass for inspiraling binaries. Relations are obtained both for the subset of equal mass configurations and for the whole set of models, which includes both equal and unequal mass configurations.

\begin{figure*}
\includegraphics[width=17cm]{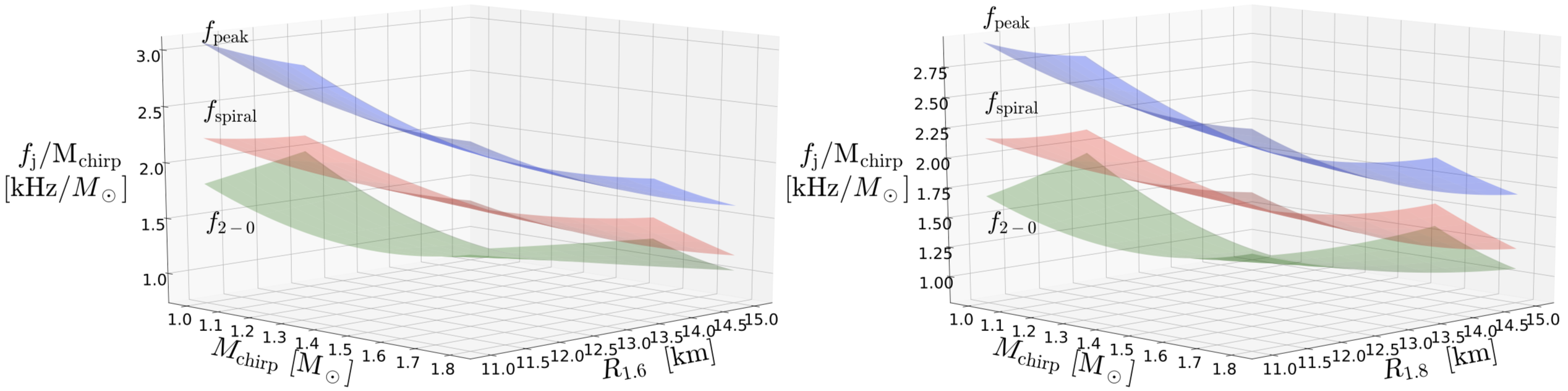}
  \caption{Empirical surfaces for frequencies with $R_{1.6}$ and $R_{1.8}$
and for all mass configurations. The blue surface corresponds to $f_{\mathrm{peak}}$
, the red surface to $f_{\mathrm{spiral}}$ and the green surface to $f_{\mathrm{2-0}}$.
The surfaces are shown only in regions where data points are available.}
  \label{fRMsurfaces2gether}
\end{figure*}


The two-parameter empirical relations of the form $f_j(R_{\rm x}, M_{\rm chirp})$ were chosen to be second-order expansions in the two parameters (including a mixed term):
\begin{equation}
  \begin{split}
    f_j / M_{\mathrm{chirp}}= b_0 + b_1 M_{\mathrm{chirp}} + b_2 R_{\mathrm{x}} + b_3 M_{\mathrm{chirp}}^2 \\
    +b_4 R_{\mathrm{x}} M_{\mathrm{chirp}}+ b_5 R_{\mathrm{x}}^2.
  \end{split}
  \label{fRM}
\end{equation}
This relation was obtained for different values of the mass of the fiducial nonrotating NS models (different values of $\rm x$ in $R_{\rm x}$). Specifically, we employ $R_{1.2}$, $R_{1.4}$, $R_{1.6}$ and $R_{1.8}$. In each case, the maximum residual and the {\it adjusted} coefficient of determination $R^2$  was evaluated (see Table \ref{table:fRM} in Appendix \ref{Appendix.E}). Below, we present for each post-merger frequency the empirical relation that has the smallest error.

\subsection{Empirical relations for $f_{\mathrm{peak}}$}

For the dominant postmerger peak frequency $f_{\rm peak}$ and using the subset of {\it equal-mass} configurations,  
the empirical relation with the smallest error is obtained for NSs of mass $1.6 M_\odot$:
\begin{equation}
  \begin{split}
    f_{\mathrm{peak}} / M_{\mathrm{chirp}}= 13.822  -0.576 M_{\mathrm{chirp}} -1.375 R_{1.6} \\ 
    + 0.479 M_{\mathrm{chirp}}^2 -0.073 R_{1.6} M_{\mathrm{chirp}}+ 0.044 R_{1.6}^2.
  \end{split}
  \label{fRM1}
\end{equation}
This fit has a maximum residual which translates to 0.196 kHz over the whole parameter space and $R^2=0.98$. The coefficients $b_0$ -- $b_5$ for the 
empirical relations constructed for other masses are shown in Table \ref{table:fRM} in Appendix B. The maximum residual in $f_{\rm peak}$ ranges from 0.196 kHz to 0.257 kHz.

For the {\it whole set} of models (including both equal and unequal masses), we display the empirical relations of the form of Eq. (\ref{fRM}), for $R_{\rm x}=$1.2,  1.4, 1.6 and 1.8 $M_\odot$, in Fig. \ref{fRMsurfaces} (notice that the surfaces in this figure are only shown in regions where data points are available, whereas for higher chirp masses and soft EOSs, i.e.\ small NS radii, the merger remnant directly collapses to a black hole~and does not produce strong postmerger GW emission - see \cite{Bauswein2013}  for an empirical relation for the threshold mass to collapse).
The empirical relation with the smallest residual is obtained for neutron stars of mass $1.8 M_\odot$: 
\begin{equation}
  \begin{split}
    f_{\mathrm{peak}} / M_{\mathrm{chirp}}= 10.942-0.369 M_{\mathrm{chirp}}-0.987 R_{1.8}  \\
    + 1.095 M_{\mathrm{chirp}}^2 -0.201 R_{1.8} M_{\mathrm{chirp}}+ 0.036 R_{1.8}^2,
  \end{split}
  \label{fRM2}
\end{equation}
which has a maximum residual that translates to 0.247 kHz over the whole parameter space and $R^2=0.976$. The coefficients $b_0$ -- $b_5$ for the 
empirical relations constructed for other masses are shown in Table \ref{table:fRM} in Appendix \ref{Appendix.B}. The maximum residual  in $f_{\rm peak}$ ranges from 0.247 kHz to 0.374 kHz.

\subsection{Empirical relations for $f_{\mathrm{2-0}}$}

For the secondary postmerger frequency $f_{2-0}$ and using the subset of {\it equal-mass} configurations,  
the empirical relation with the smallest error is obtained for neutron stars of mass $1.6 M_\odot$:
\begin{equation}
  \begin{split}
    f_{\mathrm{2-0}} / M_{\mathrm{chirp}}= 8.943 + 4.059 M_{\mathrm{chirp}}-1.332 R_{1.6}  \\
    -0.358 M_{\mathrm{chirp}}^2 -0.182 R_{1.6} M_{\mathrm{chirp}}+ 0.048 R_{1.6}^2,
 \end{split}
\end{equation}
with a maximum residual that translates to 0.229 kHz and $R^2=0.931$. The coefficients $b_0$ -- $b_5$ for the empirical relations constructed for other masses are shown in Table \ref{table:fRM} in Appendix \ref{Appendix.E}. The maximum residual ranges from 0.229 kHz to 0.366 kHz. 
For the {\it whole set} of models (including both equal and unequal masses), the empirical relation with the smallest error is obtained  for neutron stars of mass $1.6 M_\odot$: 
\begin{equation}
  \begin{split}
    f_{\mathrm{2-0}} / M_{\mathrm{chirp}}= 9.586 + 4.09 M_{\mathrm{chirp}}- 1.427 R_{1.6}  \\
    + 0.048 M_{\mathrm{chirp}}^2 -0.261 R_{1.6} M_{\mathrm{chirp}}+ 0.055 R_{1.6}^2,
 \end{split}
\end{equation}  
with a maximum residual that translates to 0.252 kHz and $R^2=0.947$.

The coefficients $b_0$ -- $b_5$ for the empirical relations constructed for other masses are shown in Table \ref{table:fRM} in Appendix \ref{Appendix.E}. The maximum residual  in $f_{\rm 2-0}$ ranges from 0.252 kHz to 0.383 kHz.

\subsection{Empirical relations for $f_{\mathrm{spiral}}$}

For the secondary postmerger frequency $f_{\rm spiral}$ and using the subset of {\it equal-mass} configurations,  
the empirical relation with the smallest error is obtained for neutron stars of mass $1.8 M_\odot$:
\begin{equation}
  \begin{split}
    f_{\mathrm{spiral}} / M_{\mathrm{chirp}}= 6.264 + 1.929 M_{\mathrm{chirp}}-0.645 R_{1.8} \\  
    + 0.881 M_{\mathrm{chirp}}^2 -0.311 R_{1.8} M_{\mathrm{chirp}}+ 0.03 R_{1.8}^2,
 \end{split}
\end{equation}
with a maximum residual that translates to 0.286 kHz and $R^2=0.944$. The coefficients $b_0$ -- $b_5$ for the empirical relations constructed for other masses are shown in  Table \ref{table:fRM} in Appendix \ref{Appendix.E}. The maximum residual in $f_{\rm spiral}$ ranges from 0.286 kHz to 0.422 kHz.

For the {\it whole set} of models (including both equal and unequal masses), the empirical relation with the smallest error is obtained again for neutron stars of mass $1.8 M_\odot$: 
\begin{equation}
  \begin{split}
    f_{\mathrm{spiral}} / M_{\mathrm{chirp}}= 5.846 + 1.75 M_{\mathrm{chirp}}-0.555 R_{1.8} \\ 
    + 1.002 M_{\mathrm{chirp}}^2 -0.316 R_{1.8} M_{\mathrm{chirp}}+ 0.026 R_{1.8}^2,
 \end{split}
\end{equation}
with a maximum residual that translates to 0.27 kHz and $R^2=0.93$ .
The coefficients $b_0$ -- $b_5$ for the empirical relations constructed for other masses are shown in  Table \ref{table:fRM} in Appendix \ref{Appendix.E}. The maximum residual in $f_{\rm spiral}$ ranges from 0.27 kHz to 0.438 kHz.

\subsection{Comparison of distinct postmerger frequencies}

In Fig. \ref{fRMsurfaces2gether}. we display the  surfaces corresponding to the empirical relations for the three different postmerger frequencies $f_{\mathrm{peak}}$, $f_{\mathrm{spiral}}$ and $f_{\mathrm{2-0}}$ for the whole CFC/SPH dataset, as a function of $M_{\rm chirp}$ and $R_{\rm x}$ (using $R_{1.6}$ in the left panel and and $R_{1.8}$ in the right panel). The surfaces are shown only in regions where data exist. It is clear that the three frequencies are \textit{distinct} in the whole parameter space. This verifies that the two secondary post-merger frequencies $f_{\mathrm{2-0}}$ and $f_{\mathrm{spiral}}$ are distinct, each satisfying a different empirical relation, as proposed in \cite{Bauswein2015}. Our findings are in contrast with the  "quasi-universal" relation that was initially proposed in \cite{Takami2014,Takami2015} for a single secondary postmerger frequency, denoted there as $f_1$. Ref. \cite{Rezzolla2016} accepts the existence of distinct postmerger frequencies, noting that their $f_1$ frequencies coincide with $f_{\rm spiral}$ in many models and with a different mode in other models (the $f_{2-0}$ frequency is identified in some models), but $f_1$ is still treated as a single feature of the post-merger spectrum that appears to satisfy a quasi-universal relation in the whole parameter space. Inspecting the data for the different extracted frequencies published in \cite{Rezzolla2016}, one can make the case that a) their $f_1$ frequency coincides with  $f_{\rm spiral}$ of \cite{Bauswein2015} in part of the parameter space, whereas it coincides with $f_{2-0}$ in other parts of the parameter space. This has already been remarked in \cite{Bauswein2015} and argued to
be fully in line with the therein devised unified picture of postmerger
GW emission. This scheme explains (by the underlying physical mechanisms)
which secondary peaks are particularly pronounced for different setups
(binary mass, EOS) and may thus be denoted as $f_1$ (see also
\cite{Clark2016,Bauswein2016,Bauswein2019} for further explanations).

Since $f_{\rm spiral}$ and $f_{2-0}$ are in fact distinct frequencies of different origin, that do not satisfy universal relations (unless one restricts to fixed masses), it follows that the quasi-universal relation for $f_1$ suggested in \cite{Takami2014,Takami2015} and again in \cite{Rezzolla2016} can only thought of as a very rough relation, having a large spread of data points (as is also evident from several outliers in the relevant figures published in the above references). The $f_{\rm spiral}$ frequency (and hence also $f_1$ in  \cite{Takami2014,Takami2015,Rezzolla2016}) is, in reality,  not universal, but satisfies relations of the form (\ref{fRM}) for each chosen mass of nonrotating models (see also Fig. 7 in \cite{Bauswein2015} regarding the non-universality of $f_{\rm spiral}$ and \cite{2019arXiv190106969B} for a more extended discussion). Furthermore, \cite{Rezzolla2016} suggest that
(in their notation) $f_2 \simeq (f_1 + f_3)/2$. But, there is  no a priori reason for this relation to hold for models where $f_1$ is in fact $f_{\rm spiral}$.   Instead, the existence of the quasi-linear combination frequencies $f_{2-0}$ and $f_{2+0}$ naturally implies  $f_2=f_{\rm peak}=(f_{2-0}+f_{2+0})/2.$

\begin{figure}
 \includegraphics[width=8.5cm]{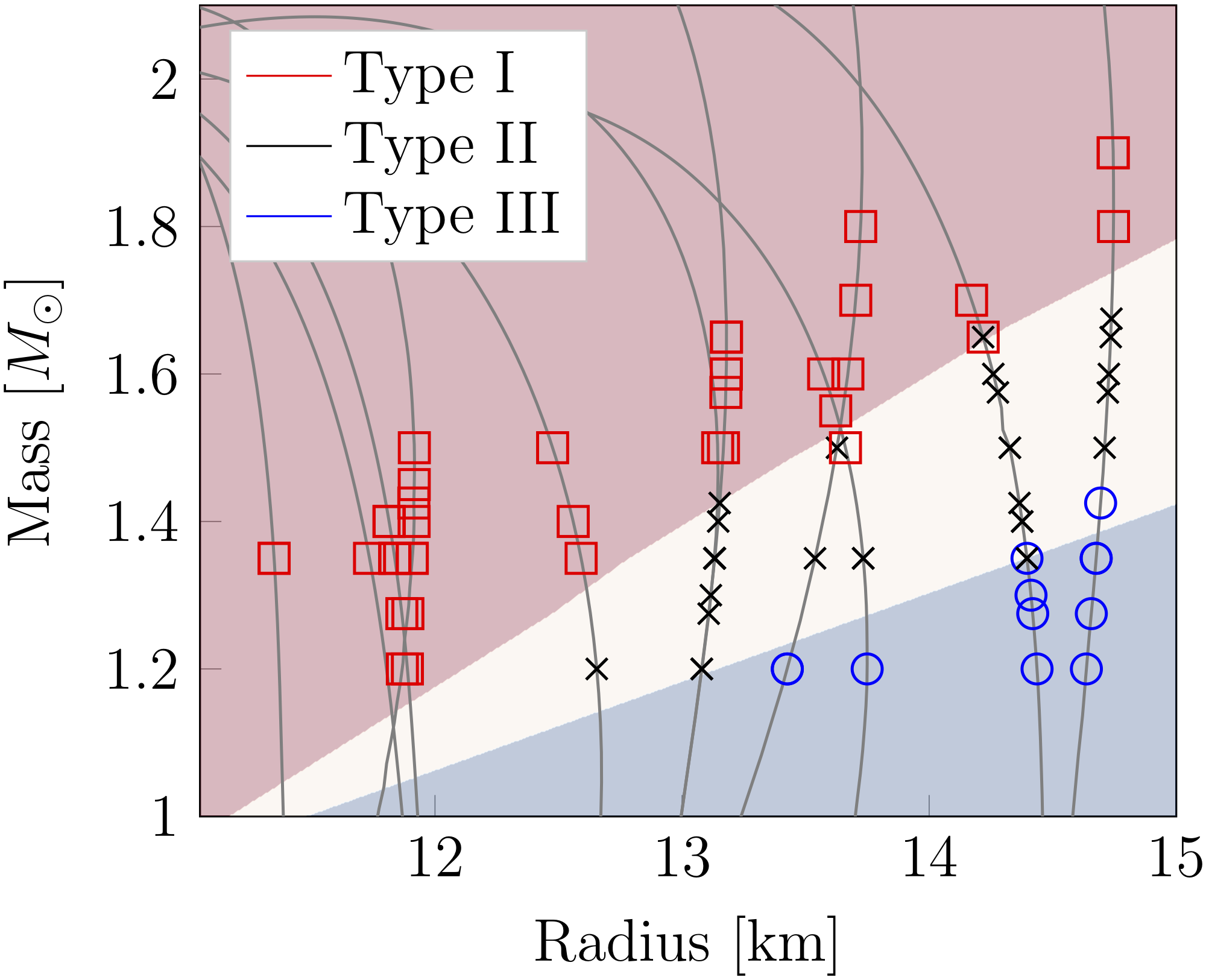}
\caption{Spectral classification of the postmerger GW emission, as obtained by a machine-learning algorithm, applied to the whole CFC/SPH data set. The classification is shown in the mass vs. radius parameter space of isolated, nonrotating neutron star models, constructed with various EOS and masses. A clustering algorithm separates the models into three different types (shown as red boxes for Type I, black $\times$ for Type II\ and blue circles for Type III). Then, a supervised-learning classification algorithm locates the borders between the three different types in this parameter space (see text for details).  The region corresponding to each type is shown in different color. The results confirm the spectral classification scheme introduced in  \cite{Bauswein2015}. Compare to Fig.~5 in~\cite{Bauswein2015}, where waveform models were classified manually, yielding a very similar pattern that is here reproduced by an automated machine-learning algorithm.}
\label{fig:classification}
\end{figure}

\section{Spectral classification of postmerger frequencies USING\ MACHINE LEARNING}
\label{sec:machine}

In \cite{Bauswein2015}, a spectral classification scheme was introduced, based on the relative amplitudes between the postmerger  $f_{\mathrm{2-0}}$ and $f_{\mathrm{spiral}}$ frequencies (see also \cite{2019arXiv190106969B} for a recent review). Here, we 
reproduce the classification of \cite{Bauswein2015}, using a machine-learning algorithm.

We choose to define the \textit{distance} between two waveforms $s$ and $h$ to be 
\begin{equation}
\mathcal{D} = 1- \mathcal{M},
\end{equation}
where $\mathcal{M}$ is the match\begin{equation}
\mathcal{M} = \underset{t_0,\phi_0}{\mathrm{max}} \frac{(s|h)}{\sqrt{(s|s)
(h|h)}},
\end{equation}
with (.|.) being the scalar product
\begin{equation}
(s|h) = 4 \mathbb{R}\mathrm{e} \int_{f_{\mathrm{low}}}^{f_{\mathrm{high}}}
\frac{\tilde{s}(f) \tilde{h}^\ast (f)}{S_n(f)}df,
\end{equation}
implemented through \cite{pycbc} (we note that for the purpose described below, other definitions of the distance between two waveforms may also be used).

Above, we denote with $\tilde{s}$ the Fourier transform of a waveform $s$ and with  $\tilde{s}^\ast$ its complex conjugate. $S_n(f)$ corresponds to the advanced LIGO BNS-optimized noise \cite{bns-psd}.

We calculated the   $n\times
n$ distance matrix between all of the $n= 89$. GW spectra of the
whole CFC/SPH dataset, in the frequency range between $f_{\mathrm{low}}=1$
kHz and $f_{\mathrm{high}} = 4$ kHz (in which the three dominant postmerger frequencies lie). 

The data were clustered with two algorithms of the
publicly available python library Scikit-Learn \cite{scikit-learn}. Both algorithms
detect the number of distinct classes (without any prior information
on their  possible number)  and depend on specific input parameters related
to their algorithmic implementation. Both the Affinity Propagation algorithm, with
a damping factor of 0.82 and a preference of 0.34 and the DBSCAN algorithm, with  parameters $\varepsilon=0.05$ and
a minimum of six points per class detected the existence of  \textit{three} distinct classes, as was proposed in \cite{Bauswein2015}. We retain the same nomenclature as  in \cite{Bauswein2015}, that is, we call a postmerger spectrum Type I when   $f_{\mathrm{2-0 }}$ is stronger than  $f_{\mathrm{spiral}}$ (occurring for soft EOS and  total binary mass not far from  the threshold mass to prompt collapse), Type II when these two secondary postmerger frequencies have comparable amplitudes  (occurring for moderately soft EOS and  intermediate total binary masses) and Type III when  $f_{\mathrm{spiral}}$ is stronger than 
 $f_{\mathrm{2-0 }}$ (occurring for stiff EOS and  total binary mass far from  the threshold
mass to prompt collapse), see  \cite{Bauswein2015,Bauswein2016,Clark2016,2019arXiv190106969B} for a more detailed description.
 
Fig. \ref{fig:classification} shows the different models of
the
whole CFC/SPH dataset, 
in a mass vs. radius graph, where in each case the mass and radius of the isolated neutron stars before merger is indicated (for each EOS that was used). In the case of unequal mass mergers, the isolated model is shown for $M_{\rm tot}/2$, where $M_{\rm tot}=M_1+M_2$ is the total mass of the individual
stars. Type I models are showns as red boxes, Type as black $\times$ and Type III as blue circles. The labels of each data point  are used in a classification algorithm, in order to find the borders between the different spectral classes in the mass vs. radius parameter space. Specifically, we used the Multi-layer Perceptron (MLP) supervised learning algorithm, with an adaptive learning rate and with the limited-memory BFGS algorithm as a
solver,  available also as part of the  Scikit-Learn library (other options were set to their default values). Fig. \ref{fig:classification}
shows the boundaries between the different spectral classes, obtained in this way (the region corresponding to each spectral
class is shown in a different color).
These results are consistent with the postmerger spectral classification scheme introduced in \cite{Bauswein2015}. 

Notice that in the case of LS375 1.8+1.8, the fundamental radial mode is around $f_0 \sim 600$Hz, which is less than the typical range for other models, because this model is very close to the threshold mass. As a result, the secondary peaks in the postmerger spectrum appear in opposite order, compared to lower-mass cases (for somewhat higher central density the remnant would have a quasi-radial frequency even smaller, tending to zero, which marks the onset of collapse). Because of this exceptional morphology, the spectrum of this model was classified as type II by the algorithm described above. Demanding that the  fundamental radial mode frequency $f_0$ only decreases as one approaches the threshold mass to collapse, for a given EOS, restores the correct identification of the secondary peaks and is used as an additional criterion in setting the right labels in Knowing the reason, we still show this single data point as type I in Fig. \ref{fig:classification} (in our sample this re-labeling was needed only for the  LS375 1.8+1.8 model).

\begin{figure*}[ht]
 \includegraphics[width=17cm]{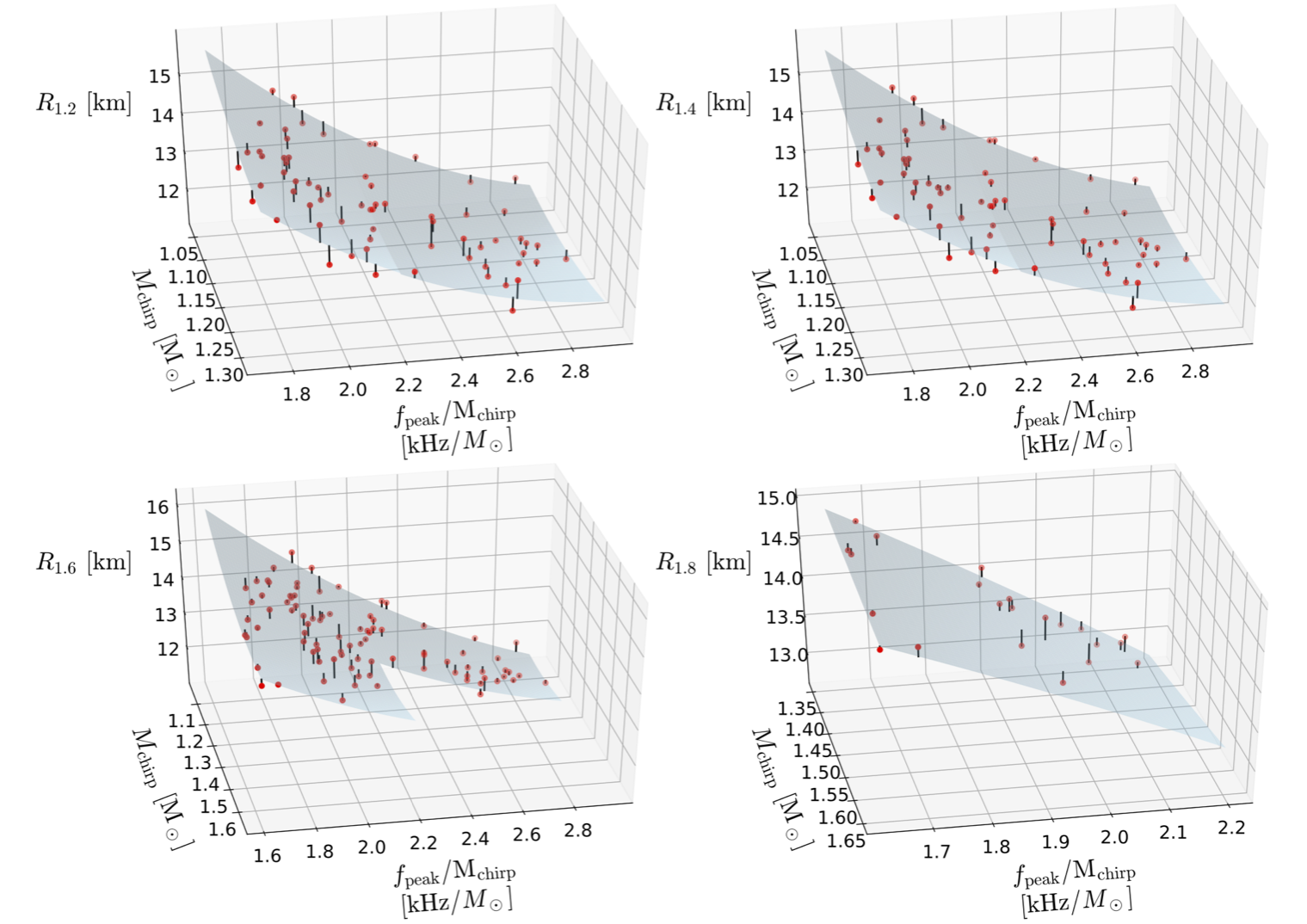}
\caption{Surfaces $R_{\mathrm{x}}(f_{\mathrm{peak}},M_{\mathrm{chirp}})$
using the whole SPH/CFC data set. Red dots correspond to simulation data
($f_{\mathrm{peak}},M_{\mathrm{chirp}})$ with the vertical axis corresponding
to the radius $R_{\rm x}$ of a nonrotating model with the same EOS as used
in each simulation (in the different
panels, the radius of nonrotating
neutron stars of mass 1.2, 1.4, 1.6 and 1.8$M_\odot$ was used).  The light
blue surfaces represent the empirical relations
of the form of Eq. (\ref{RfM}). The surfaces
are shown only in regions where data points are available.}
\label{fig:RfMsurfaces}
\end{figure*}

\begin{figure*}[ht]
 \includegraphics[width=17cm]{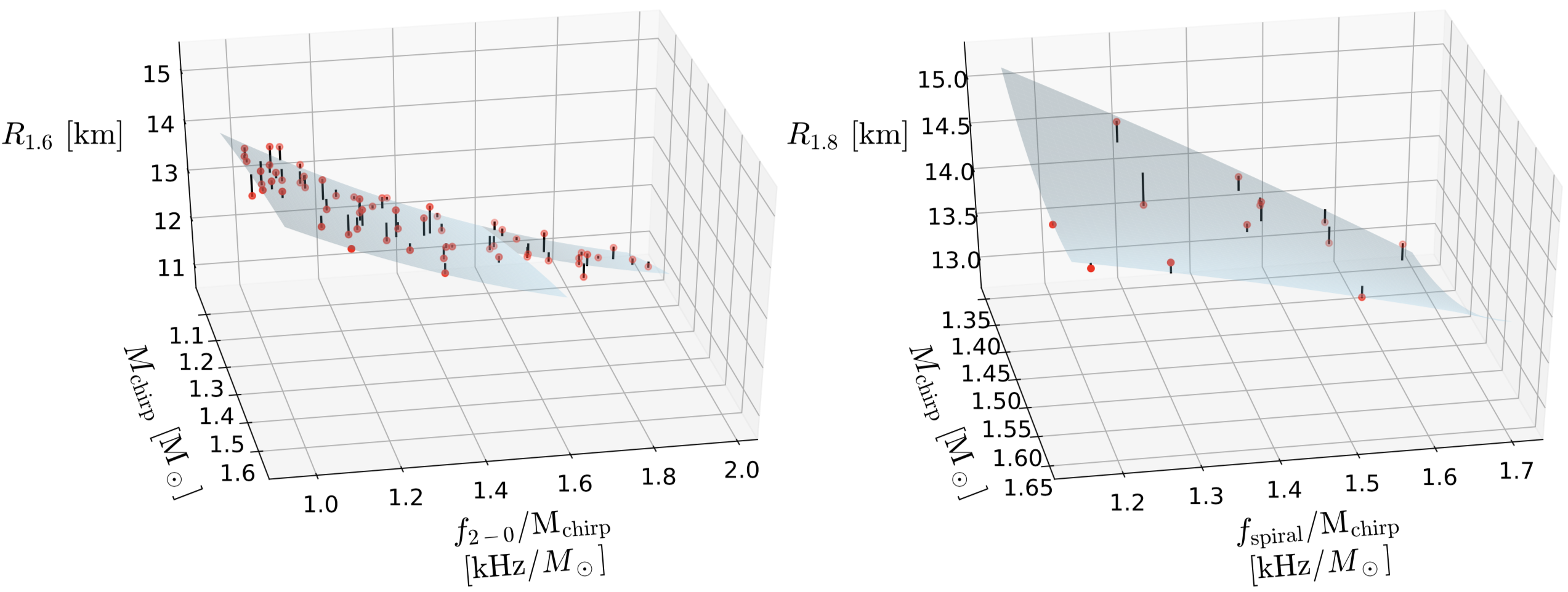}
  \caption{Surfaces $R_{\mathrm{1.6}}(f_{\mathrm{2-0}},M_{\mathrm{chirp}})$
(left panel) and  $R_{\mathrm{1.8}}(f_{\mathrm{spiral}},M_{\mathrm{chirp}})$
(right panel) using the whole SPH/CFC data set. Red dots correspond to simulation data
($f_{\mathrm{peak}},M_{\mathrm{chirp}})$ with the vertical axis corresponding
to the radius $R_{\rm 1.6}$ (or $R_{\rm 1.8}$, correspondingly) of a nonrotating model with the same EOS as used
in each simulation.  The light
blue surfaces represent the empirical relations
of the form of Eq. (\ref{RfM}). The surfaces
are shown only in regions where data points are available.\\ }
  \label{fig:fRMsurfaces2}
\end{figure*}

\section{Empirical relations for radii based on the Bauswein et al.
CFC/SPH catalogue}\label{sec:rad}
The empirical relations for postmerger frequencies as function of radius and chirp mass,  of the form $f_j(R_{\rm x}, M_{\rm
chirp})$  investigated in Sec. \ref{sec:freq}, can be inverted, in order to obtain relations for chosen radii or nonrotating models  as function of postmerger frequencies and chirp mass, of the form $R_{\rm x}(f_j, M_{\rm chirp})$, where $\rm x$ can be $\{1.4, 1.6, 1.8\}$ and $j = \{ \mathrm{peak}, \mathrm{spiral}, \mathrm{2-0}  \}$. Instead of direct inversion of the empirical relations found in Sec. \ref{sec:freq}, we construct new relations applying a least-squares minimization to  the same data. After investigating different possible forms, we found that a good choice is the second order expansion in both $f_j$ and $M_{\rm
chirp}$ (including the mixed term)
\begin{equation}
  \begin{split}
    R_{\mathrm{x}}= b_0 + b_1 M_{\mathrm{chirp}} + b_2 f_{\mathrm{j}}/M_{\mathrm{chirp}}
+b_3 M_{\mathrm{chirp}}^2\\ +b_4 f_{\mathrm{j}} +b_5 \left( f_{\mathrm{j}}/M_{\mathrm{chirp}}
\right)^2,
    \end{split}
  \label{RfM}
\end{equation}
(more details on the performance of the above and of other investigated forms are given in Appendix \ref{Appendix.E}).
\par \par When constructing the empirical relations of the form (\ref{RfM}), we noticed the following optimization: for the $R_{1.2}(f_{\mathrm{j}},M_{\mathrm{chirp}})$
and $R_{1.4}(f_{\mathrm{j}},M_{\mathrm{chirp}})$ relations, we use only the
data for which $M_{\mathrm{chirp}} < 1.3$, whereas for the  $R_{1.8}(f_{\mathrm{j}},M_{\mathrm{chirp}})$ we use only the data for which  $M_{\mathrm{chirp}}> 1.3$. This is natural, since the lower mass ($M_{\mathrm{chirp}}$) binaries
are not suitable for  inferring information for neutron stars of large mass and vice versa. Since, in this way, the dataset is separated into two regions, depending on the target radius, we use  the superscript (\textless \ or \textgreater ) in naming the empirical relations. We note that for the  $R_{1.6}(f_{\mathrm{j}},M_{\mathrm{chirp}})$ relation we use the whole dataset, since this is an intermediate case.

We emphasize that in principle one should consider distinct relations for relatively small ranges in $M_{\mathrm{chirp}}$, which can be measured with high precision, as those relations should yield the tightest correlations and thus the smallest errors in radius measurements through postmerger GW emission. This approach, however, requires an even larger set of simulations with systematically varied binary mass parameters, especially the mass ratio.

\subsection{Empirical relations for ${R_{1.2}}$}

For ${R_{1.2}}$ and using the subset of {\it equal-mass} configurations,  
the empirical relation with the smallest error is
\begin{equation}
  \begin{split}
    R_{1.2}^{<}= 52.201 -29.769 M_{\mathrm{chirp}} -15.398 f_{\mathrm{peak}}/M_{\mathrm{chirp}}\\ +8.918 M_{\mathrm{chirp}}^2 +3.333 f_{\mathrm{peak}} +1.832 \left( f_{\mathrm{peak}}/M_{\mathrm{chirp}} \right)^2,
  \end{split}
  \label{RfM_R12_e1}
\end{equation}
with a maximum residual of 0.52 km and $R^2=0.945$. The coefficients $b_0$ -- $b_5$ for the 
empirical relations constructed when using other frequencies are shown in Table \ref{table:RfM} in Appendix \ref{Appendix.E}. The maximum residual ranges between 0.52 km and 0.8 km.
 
For the {\it whole set} of models (including both equal and unequal masses), the empirical relation with the smallest error is \begin{equation}
  \begin{split}
    R_{1.2}^{<}= 56.906 -37.252 M_{\mathrm{chirp}} -15.701 f_{\mathrm{peak}}/M_{\mathrm{chirp}}\\ +11.756 M_{\mathrm{chirp}}^2 +3.638 f_{\mathrm{peak}} +1.83 \left( f_{\mathrm{peak}}/M_{\mathrm{chirp}} \right)^2,
  \end{split}
  \label{RfM_R12_e0}
\end{equation}
with a maximum residual of 0.526 km and $R^2=0.951$.
The coeffients $b_0$ -- $b_5$ for the empirical relations constructed for other frequencies are shown in Table \ref{table:RfM} in Appendix \ref{Appendix.E}. The maximum residual ranges between 0.526 km and 0.737 km.

\subsection{Empirical relations for ${R_{1.4}}$}
For  ${R_{1.4}}$ and using the subset of {\it equal-mass} configurations, the empirical relation with the smallest error is 
\begin{equation}
  \begin{split}
    R_{1.4}^{<}=51.229 -30.463 M_{\mathrm{chirp}} -14.143 f_{\mathrm{peak}}/M_{\mathrm{chirp}}\\ +9.46 M_{\mathrm{chirp}}^2 + 3.09 f_{\mathrm{peak}} +1.612 \left( f_{\mathrm{peak}}/M_{\mathrm{chirp}} \right)^2,
  \end{split}
  \label{RfM_R14_e1}
\end{equation}
with a maximum residual of 0.412 km and $R^2=0.966$. The coefficients $b_0$ -- $b_5$ for the 
empirical relations constructed when using other frequencies are shown in Table \ref{table:RfM} in Appendix \ref{Appendix.E}. The maximum residual ranges between 0.412 km and 0.731 km.
 
For the {\it whole set} of models (including both equal and unequal masses), the empirical relation with the smallest error is 
\begin{equation}
  \begin{split}
    R_{1.4}^{<}=55.809 -37.642 M_{\mathrm{chirp}} -14.473 f_{\mathrm{peak}}/M_{\mathrm{chirp}}\\ +12.15 M_{\mathrm{chirp}}^2 +3.41 f_{\mathrm{peak}} +1.609 \left( f_{\mathrm{peak}}/M_{\mathrm{chirp}} \right)^2,
  \end{split}
  \label{RfM_R14_e0}
\end{equation}
with a maximum residual of 0.493 km and $R^2=0.968$.
The coefficients $b_0$ -- $b_5$ for the empirical relations constructed when using other frequencies are shown in Table \ref{table:RfM} in Appendix \ref{Appendix.E}. in Appendix B. The maximum residual ranges between 0.493 km and 0.676 km.

\subsection{Empirical relations for ${R_{1.6}}$}
For  ${R_{1.6}}$ and using the subset of {\it equal-mass} configurations,  
the empirical relation with the smallest error is obtained when using the dominant postmerger frequency $f_{\mathrm{peak}}$ 

\begin{equation}
  \begin{split}
    R_{1.6}= 41.316 - 16.654 M_{\mathrm{chirp}} -12.458 f_{\mathrm{peak}}/M_{\mathrm{chirp}}\\ +3.722 M_{\mathrm{chirp}}^2 +2.936 f_{\mathrm{peak}} +1.269 \left( f_{\mathrm{peak}}/M_{\mathrm{chirp}} \right)^2,
  \end{split}
  \label{RfM_R16_e1}
\end{equation}
with a maximum residual of 0.462 km and $R^2=0.97$. A comparable performance is obtained when using the secondary postmerger
frequency $f_{2-0}$ 
\begin{equation}
  \begin{split}
    R_{1.6}= 15.271 + 4.123 M_{\mathrm{chirp}} -6.661 f_{\mathrm{2-0}}/M_{\mathrm{chirp}}\\
-1.188 M_{\mathrm{chirp}}^2 +1.23 f_{\mathrm{2-0}} +0.783 \left( f_{\mathrm{2-0}}/M_{\mathrm{chirp}}
\right)^2,
 \end{split}
\end{equation}
which has a maximum residual
of 0.465 km and $R^2=0.942$. The coefficients $b_0$ -- $b_5$ for the 
empirical relation constructed when using $f_{\rm spiral}$ are shown in Table \ref{table:RfM} in Appendix \ref{Appendix.E}. Among all different choices, the maximum residual ranges between 0.462 km and 0.706 km. We stress that secondary peaks being weaker in gravitational waves are more difficult to detect and typically have a larger full width at half maximum (FWHM) implying that the error of a frequency measurement of secondary features in a GW detection will be larger compared to that of the main peak.
 
For the {\it whole set} of models (including both equal and unequal masses), the empirical relation with the smallest error is obtained when using the secondary postmerger frequency $f_{2-0}$ 
\begin{equation}
  \begin{split}
    R_{1.6}= 17.764 + 2.497 M_{\mathrm{chirp}} -8.797 f_{\mathrm{2-0}}/M_{\mathrm{chirp}}\\
-0.639 M_{\mathrm{chirp}}^2 +1.393 f_{\mathrm{2-0}} +1.452 \left( f_{\mathrm{2-0}}/M_{\mathrm{chirp}}
\right)^2,
 \end{split}
\end{equation}
with a maximum residual of 0.518 km and $R^2=0.955$. A comparable performance
is obtained when using the dominant postmerger
frequency  $f_{\mathrm{peak}}$ 
\begin{equation}
  \begin{split}
    R_{1.6}= 43.796 -19.984 M_{\mathrm{chirp}} -12.921 f_{\mathrm{peak}}/M_{\mathrm{chirp}}\\
+4.674 M_{\mathrm{chirp}}^2 +3.371 f_{\mathrm{peak}} +1.26 \left( f_{\mathrm{peak}}/M_{\mathrm{chirp}}
\right)^2,
  \end{split}
  \label{RfM_R16_e0}
\end{equation}
with a maximum residual of 0.526 km and $R^2=0.969$. 
The coefficients $b_0$ -- $b_5$ for the empirical relation constructed when using $f_{\rm spiral}$ is shown in Table \ref{table:RfM} in Appendix \ref{Appendix.E}. Among all different choices, the maximum residual ranges between 0.518 km and 0.674 km.

\begin{figure*}
  \includegraphics[width=17cm]{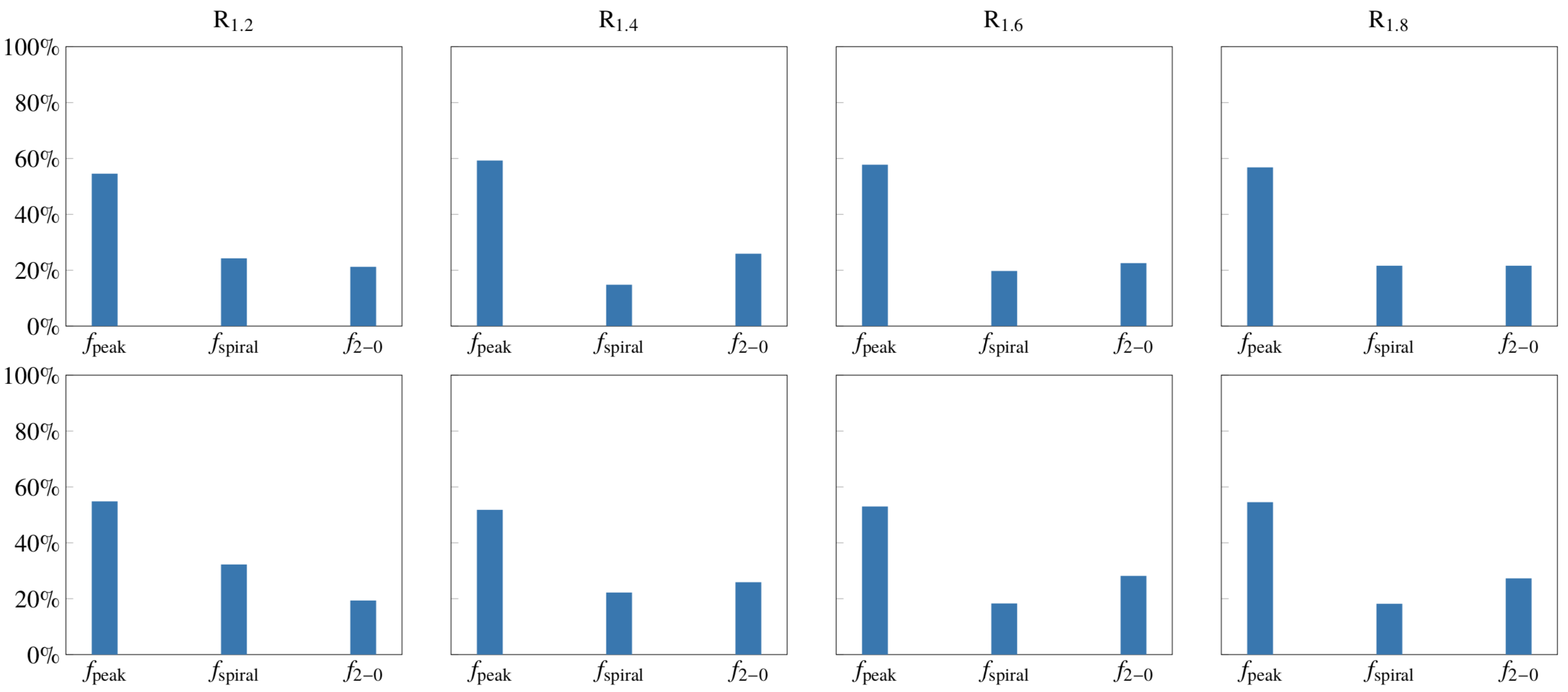}
  \caption{In each panel, the percentage of data points that are closer in
radius to each of the empirical
relations (constructed with the corresponding frequency) is shown. The top
row shows results
of equal-mass configurations only and the bottom row uses all CFC/SPH data
(see text for more explanations). Generally, these figures imply that other
statistical measures for the quality of empirical relations (involving some
sort of weighting like the 2-norm) would reveal tighter relations for $f_\mathrm{peak}$
in comparison to the subdominant frequencies.}
  \label{hist}
\end{figure*}

\subsection{Empirical relations for $\mathrm{R_{1.8}}$}
For  $\mathrm{R_{1.8}}$ and using the subset of {\it equal-mass} configurations,  
the empirical relation with the smallest error is obtained for the secondary postmerger frequency $f_{\mathrm{spiral}}$
\begin{equation}
  \begin{split}
    R_{1.8}^{>}= 55.934 -37.162 M_{\mathrm{chirp}} - 17.139 f_{\mathrm{spiral}}/M_{\mathrm{chirp}}\\ +7.961 M_{\mathrm{chirp}}^2 +9.897 f_{\mathrm{spiral}} -0.382 \left( f_{\mathrm{spiral}}/M_{\mathrm{chirp}} \right)^2,
 \end{split}
 \label{RfM_R18_e1s}
\end{equation}
with a maximum residual of 0.212 km and $R^2=0.951$. A comparable performance
is obtained when using the dominant postmerger
frequency  $f_{\mathrm{peak}}$
\begin{equation}
  \begin{split}
    R_{1.8}^{>}= 33.802 -3.069 M_{\mathrm{chirp}} -15.522 f_{\mathrm{peak}}/M_{\mathrm{chirp}}\\ -1.439 M_{\mathrm{chirp}}^2 + 4.112 f_{\mathrm{peak}} + 1.605 \left( f_{\mathrm{peak}}/M_{\mathrm{chirp}} \right)^2,
  \end{split}
  \label{RfM_R18_e1}
\end{equation} 
with a maximum residual of 0.276 km and $R^2=0.951$.  The coefficients $b_0$ -- $b_5$ for the 
empirical relation constructed when using $f_{\rm 2-0}$ are shown in Table \ref{table:RfM} in Appendix \ref{Appendix.E}. in Appendix B. Among all different choices, the maximum residual ranges between 0.212 km and 0.597 km.

For the {\it whole set} of models (including both equal and unequal masses), the empirical relation with the smallest error is \begin{equation}
  \begin{split}
    R_{1.8}^{>}= 54.467 -38.851 M_{\mathrm{chirp}} -13.992 f_{\mathrm{peak}}/M_{\mathrm{chirp}}\\ +9.305 M_{\mathrm{chirp}}^2 +8.453 f_{\mathrm{peak}} -0.614 \left( f_{\mathrm{peak}}/M_{\mathrm{chirp}} \right)^2,
  \end{split}
  \label{RfM_R18_e0}
\end{equation}
with a maximum residual of 0.275 km and $R^2=0.958$.
The coefficients $b_0$ -- $b_5$ for the empirical relations constructed for other frequencies are shown in Table \ref{table:RfM} in Appendix \ref{Appendix.E}. The maximum residual ranges from 0.275 km to 0.569 km.

\subsection{Comparing the performance of empirical relations for radii}

For the {\it whole set} of models (including both equal and unequal masses),
we display the empirical relations of the form of Eq. (\ref{RfM}), for $R_X=$1.2,
 1.4, 1.6 and 1.8 $M_\odot$, when using $f_{\rm peak}$, in Fig. \ref{fig:RfMsurfaces} (notice that the surfaces
in the different panels of this figure are only shown in  regions where data points are available). Each $R_{\rm x}$ depends mainly on $f_{\rm peak}$ and
to a smaller degree on $M_{\rm chirp}$, as anticipated from the previous results by \cite{Bauswein2012a} (see e.g.~\cite{2019arXiv190106969B} for a review).
If one would not be interested in the smallest possible residual, then a linear approximation (a plane surface in this figure) would be sufficient. But, for high accuracy, the extension to second order, as is done here through Eq. (\ref{RfM}), is required.

Since the empirical relations $R_{\mathrm{1.6}}(f_{\mathrm{2-0}},M_{\mathrm{chirp}})$
 and  $R_{\mathrm{1.8}}(f_{\mathrm{spiral}},M_{\mathrm{chirp}})$ had
a comparable accuracy to the corresponding relations with $f_{\rm peak}$, we display these in Fig. \ref{fig:fRMsurfaces2}. For $R_{\mathrm{1.6}}(f_{\mathrm{2-0}},M_{\mathrm{chirp}})$, the dependence on $M_{\rm chirp}$ is weak, but for $R_{\mathrm{1.8}}(f_{\mathrm{spiral}},M_{\mathrm{chirp}})$ it is strong in the limit of low masses. 

\begin{figure*}
  \includegraphics[width=15cm]{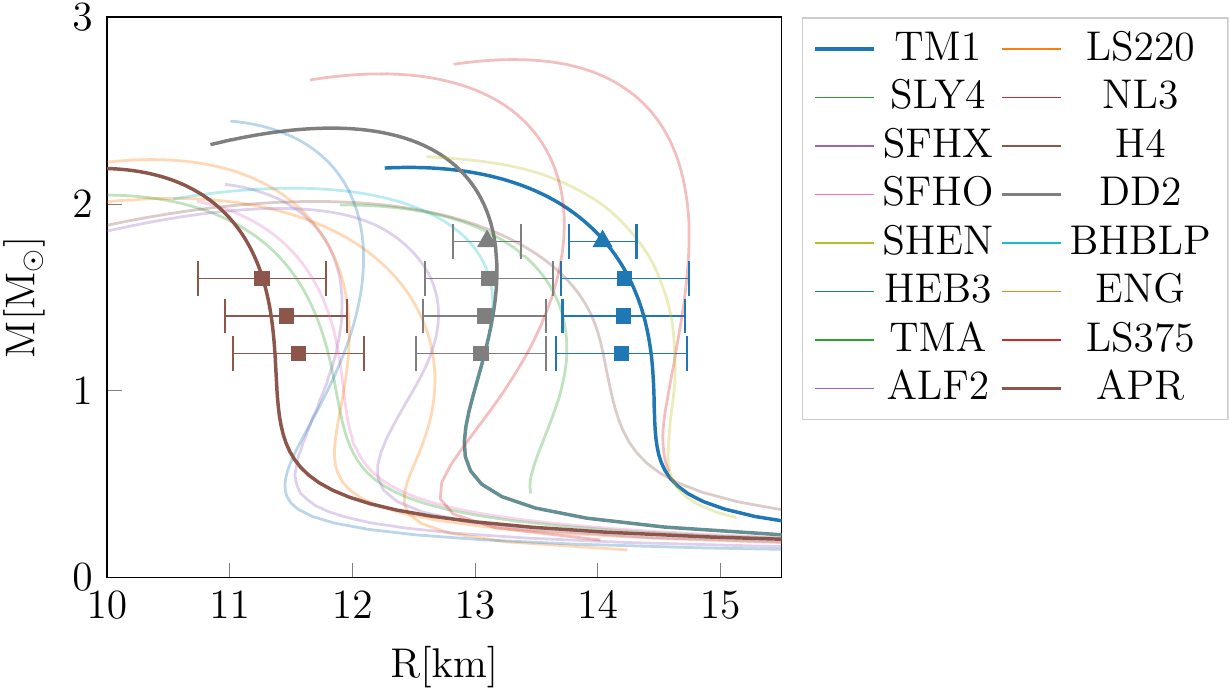}
  \caption{Predictions for radius determinations at various masses using
$f_{\mathrm{peak}}$ in the empirical relations
(\ref{RfM}) assuming mergers with either $1.35+1.35 M_\odot$\ (squares) or
 $1.6+1.6 M_\odot$ (triangles), for three different candidate EOS.  We assume
that either the APR, DD2 or TM1 EOS is the correct
EOS of high-density matter and predict the radius for certain masses.  In
the mass range of $1.2-1.6 M_\odot$, the true radius is within the maximum
possible residual of $\sim \pm0.5$km from the predicted radius. For $1.8
M_\odot$
(EOS DD2 and TM1 only) the true radius is within a smaller maximum
possible residual of  only $\sim \pm0.28$km from the predicted radius. }
\label{fig:tovplot}
\end{figure*}
 
Relations of the form of Eq. (\ref{RfM}) 
can be used to obtain the radii \(R_{\rm x}\)
at different masses, when using any of the three postmerger frequencies $f_{\rm peak}, f_{2-0}$ or $f_{\rm spiral}$. 
We investigated the performance of each empirical relation
in obtaining \(R_{\rm x}\)
 and a comparison is shown in Fig. \ref{hist} (the top row corresponds to equal-mass models only). In each
panel, we show the percentage of data points that have the smallest residual among the different choices for the postmerger frequency (each column corresponds to a different mass   \(R_{\rm x}\)).  For all different masses, the corresponding radius of nonrotating stars is obtained more accurately when using the empirical
relations for $f_{\mathrm{peak}}$ in more than 50\% of cases. For the remaining cases, the empirical relations using either the  $f_{\mathrm{2-0}}$
  or the  $f_{\mathrm{spiral}}$  frequencies were more accurate in predicting radii, with  the relations using $f_{\mathrm{2-0}}$
  outperforming  the relations using $f_{\mathrm{spiral}}$  for most masses,
except for the lowest mass of \(1.2 M_{\odot }\)
maybe to help to explain these data:
These statistics exemplify that for the majority of all models the $f_\mathrm{peak}$
data points are closest to the respective empirical relation, whereas the
data points of secondary peaks show a much larger scatter on average. Generally, these figures imply that other
statistical measures for the quality of empirical relations (involving some
sort of weighting like the 2-norm) would reveal tighter relations for $f_\mathrm{peak}$
in comparison to the subdominant frequencies, but as commented in
Sect. \ref{sec:sum} we do not follow this approach here.  

We emphasize that the errors we quote for radius measurements through relations of the form  of Eq. (\ref{RfM})  represent \textit{upper limits} (the maximum residuals correspond to the worst case in the whole sample) using our currently large set of representative EOS. These maximum residuals can improve in two ways: First, in an actual detection, binary mass parameters, such as  the chirp mass and the mass ratio, will be measured. Hence, employing\textit{ optimized} relations that can be constructed for a narrower range of measured binary parameters will likely result in significantly smaller residuals. Second, future EOS constraints from a variety of experimental and observational methods may faithfully restrict the sample of representative EOS to a smaller sample, spanning a narrower region in the mass vs. radius parameter space. We therefore anticipate that our empirical relations of the form  of Eq. (\ref{RfM})  will significantly improve over time. 

In a realistic detection scenario, the signal-to-noise ratio (SNR) of each frequency peak will determine its detectability and greatly influence its accuracy
in measuring radii. In this sense, we expect the dominant postmerger 
frequency $f_{\mathrm{peak}}$ to play the dominant role  in measuring radii, with the other two frequencies (typically having smaller SNR and larger width than $f_{\mathrm{peak}}$) being useful for extracting additional information on the characteristics of the postmerger remnant. These considerations and the data displayed in Fig.~\ref{hist} demonstrate that  $f_{\mathrm{peak}}$ is the most promising feature for EOS constraints from the postmerger phase.

It is fortunate that the empirical relations for $R^{\rm >}_{1.8}$ have very small residuals, between 0.212 km and 0.275 km.  When one considers  the currently available observed sample of neutron stars in binary systems, it is reasonable to expect that neutron stars with a mass of 1.8$M_\odot$ will only rarely be members of merging binary systems (see e.g.~\cite{2012ApJ...757...55O,2016ARA&A..54..401O,2019ApJ...876...18F}). Even less frequent would be a case
of equal mass mergers with both stars having such a high mass. This implies that it will be quite difficult to accurately measure the radius or tidal deformability of high-mass neutron stars, when using methods based on the inspiral part of the gravitational-wave emission, i.e.\ methods based on measuring the tidal deformability (see, e.g.~\cite{2019arXiv190708534B} for a recent review and references therein) or frequencies excited through resonances (see 
e.g.~\cite{2019arXiv190500818S}). Moreover, finite-size effects decrease for higher masses as the tidal deformability is smaller. Hence, even if the inspiral of a high-mass binary is observed, the extraction of NS parameters may be more challenging and associated with larger errors. 
In contrast, the postmerger empirical relations (\ref{RfM_R18_e1s}), (\ref{RfM_R18_e1}) and  (\ref{RfM_R18_e0}) provide a competitive method for measuring the radius of high-mass neutron stars and thus for constraining the very high density part of the EOS.

\section{Constraining the mass-radius relation}\label{sec:mr}

We consider three particular case studies, where we a assume that a certain EOS is the correct one, a soft EOS, APR, an intermediate  EOS, DD2 and a stiff EOS, TM1.  For the soft EOS APR we assume that the dominant postmerger frequency $f_{\rm peak}$ is detected in a single event with  $M_{\rm chirp}<1.3M_\odot$ (specifically, from a $1.35+1.35 M_\odot$\ merger), whereas for the other two EOS we assume that  $f_{\rm peak}$ is detected in two distinct binary neutron star merger events, one with $M_{\rm chirp}<1.3M_\odot$ (a $1.35+1.35 M_\odot$\ merger) and a second with $M_{\rm chirp}>1.3M_\odot $ (a $1.6+1.6 M_\odot$\ merger).
Fig. \ref{fig:tovplot} shows the predicted radii $R_{1.2}, R_{1.4}, R_{1.6}$ and $R_{1.8}$ (the latter only for the intermediate and stiff EOS) in a mass vs. radius diagram, where also different sample EOS are shown. For each predicted radius, we show error bars that correspond to the \textit{maximum residual }  
 of each empirical relation that was used. Filled boxes correspond to empirical relations that are valid for $M_{\rm chirp}<1.3M_\odot$, while filled triangles correspond to  empirical
relations that are valid for $M_{\rm chirp}>1.3M_\odot$. 

From the results displayed in the figure, it is apparent that
our empirical relations can be used to constrain the mass-radius
relation of nonrotating neutron stars with a maximum uncertainty of
about $\pm 0.5$km in the range $1.2-1.6M_\odot$\ and with an even smaller maximum uncertainty of $\pm0.28$km for neutron stars of mass $1.8M_\odot$ (see Sec.~\ref{sec:sum} for discussion). Such radius constraints can readily be translated to constraints on the pressure vs. energy density, $P(\epsilon)$, relation, i.e.\ the EOS  (see e.g.~\cite{Fattoyev2017,Raithel2018,PhysRevLett.121.161101,2019arXiv190511212T,2019arXiv190605978C}).
 In our examples the actual recovery of the radii for individual models is much better than indicated by the error bars. This is because we assign the maximum residual as error bar because one cannot know a priori how well the true EOS of NSs follows the empirical relations. By considering a large representative sample of candidate EOS, we expect that the maximum residual among all viable EOS models provides a safe proxy for the error although it is  quite possible that the actual error will be smaller. We
emphasize again that the error can be further reduced by considering empirical relations for a fixed chirp mass or a chirp mass within a small range (recall that the chirp mass can be measured very precisely from the inspiral phase). The situation depicted in Fig.~\ref{fig:tovplot} thus represents a \textit{worst-case scenario}.

\begin{figure}[!]
  \includegraphics[width=8.5cm]{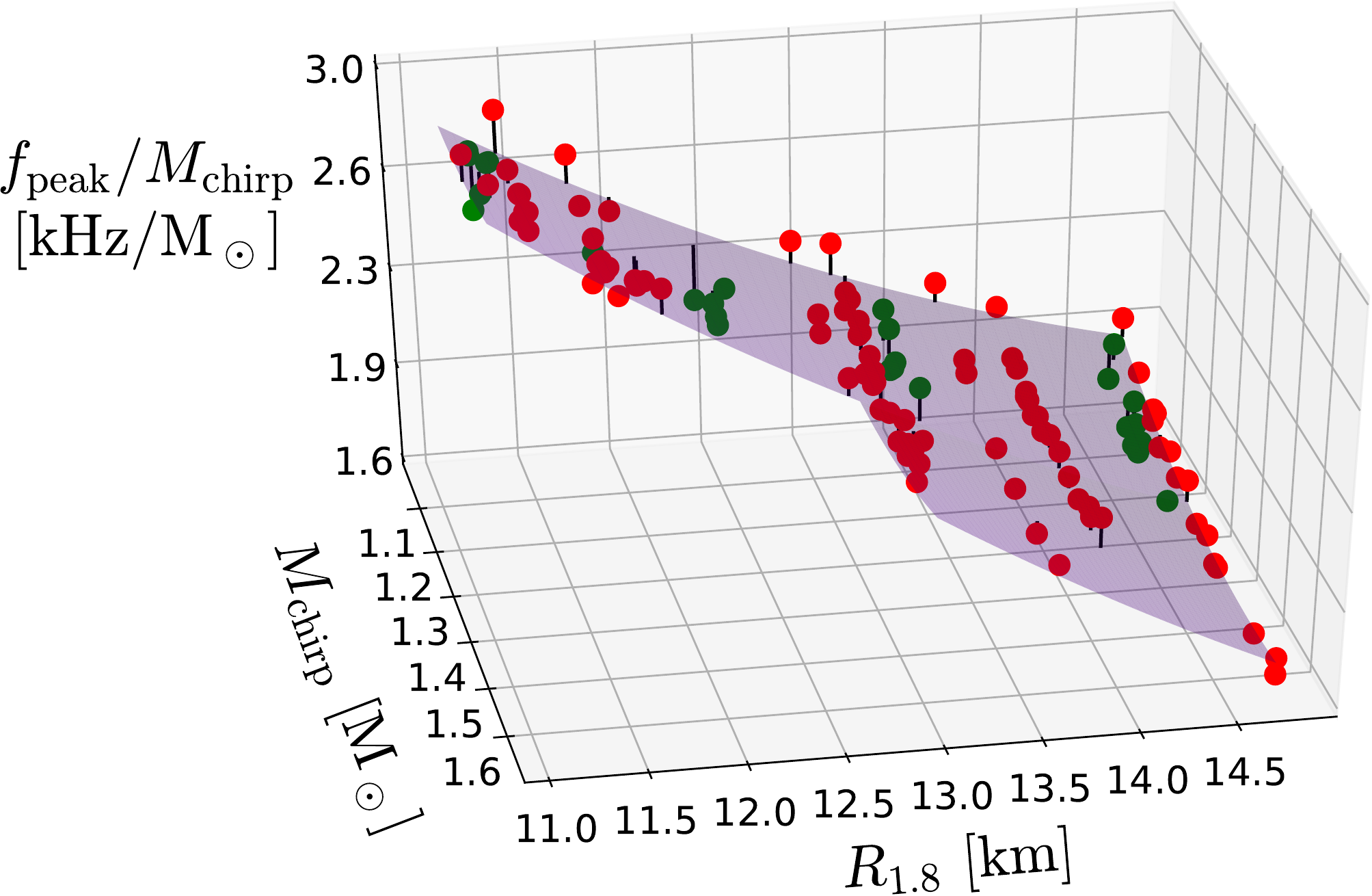}
\caption{Combined data sets surfaces for frequencies. Red points correspond
to CFC/SPH data and green points correspond to data extracted from the CoRe
GW catalogue.  surface for $R_{1.8}$ and all binary mass configurations.}
\label{Combined_fRM}
\end{figure}



\section{Validation of empirical relations using frequencies extracted from the CORE GW catalogue}
\label{CORE:relations}
Using the CoRe\ GW catalogue, we extracted the peak post-merger frequency  $f_{\mathrm{peak}}$ for each waveform and then constructed empirical relations of the form  $f_{\mathrm{peak}}(R_{\mathrm{x}},M_{\mathrm{chirp}})$ and $R_{\mathrm{x}}(f_{\mathrm{peak}},M_{\mathrm{chirp}})$  (additional relations based on other post-merger frequencies will not be reported here). The aim was to validate the empirical relations constructed with the CFC/SPH dataset of Bauswein et al. using a dataset that was obtained with very different numerical methods. The second-order dependence of the empirical relations on the dependent variables is rather weak. The CFC/SPH dataset of Bauswein et al.  had a sufficient number of data points such that second-order empirical relations lead to advantages compared to simpler first-order ones. The models of the CoRe dataset used here are fewer and the maximum residual is comparable between the choices of first-order or second-order empirical relations.
In the following, we will present some examples of second-order empirical relations constructed using the combined data sets (adding the models of the CFC/SPH and CoRe datasets).

We construct new empirical relations for the \textit{combined dataset }of the CFC/SPH models and our subset of CoRe models. For the dominant postmerger frequency peak and using the subset of {\it equal-mass}
configurations, the empirical relation with the smallest error is obtained
for neutron stars of mass $1.8 M_\odot$. 
\begin{equation}
  \begin{split}
    f_{\mathrm{peak}} / M_{\mathrm{chirp}}= 11.476  +0.025 M_{\mathrm{chirp}}
-1.102 R_{1.8} \\ 
    + 1.181 M_{\mathrm{chirp}}^2 -0.242 R_{1.8} M_{\mathrm{chirp}}+ 0.042
R_{1.8}^2,
  \end{split}
  \label{Combined_fRM_R18_e1}
\end{equation}
with a maximum residual of 0.14 kHz and $R^2 = 0.975$.
In this case, the addition of the CoRe data to the CFC/SPH dataset improves the 
empirical fit somewhat, resulting in a slightly higher $R^2$ and somewhat smaller
maximum residual than for the CFC/SPH dataset alone. 

Similarly, when using the {\it whole set} of models, the empirical relation with the smallest
error is obtained for neutron stars of mass $1.8 M_\odot$. 
\begin{equation}
  \begin{split}
    f_{\mathrm{peak}} / M_{\mathrm{chirp}}= 9.044  +0.713 M_{\mathrm{chirp}}
-0.804 R_{1.8} \\ 
    + 1.017 M_{\mathrm{chirp}}^2 -0.259 R_{1.8} M_{\mathrm{chirp}}+ 0.031
R_{1.8}^2,
  \end{split}
    \label{Combined_fRM_R18_e0}
\end{equation}
with a maximum residual of 0.197 kHz and $R^2 = 0.966$.
Figure (\ref{Combined_fRM}) shows the above empirical fit as a surface as well as the CFC/SPH data points (red dots) and the CoRe data points (green points). The distribution of the CoRe data points is in excellent agreement with the distribution of the CFC/SPH data points.

\begin{figure}[!]
  \includegraphics[width=8.5cm]{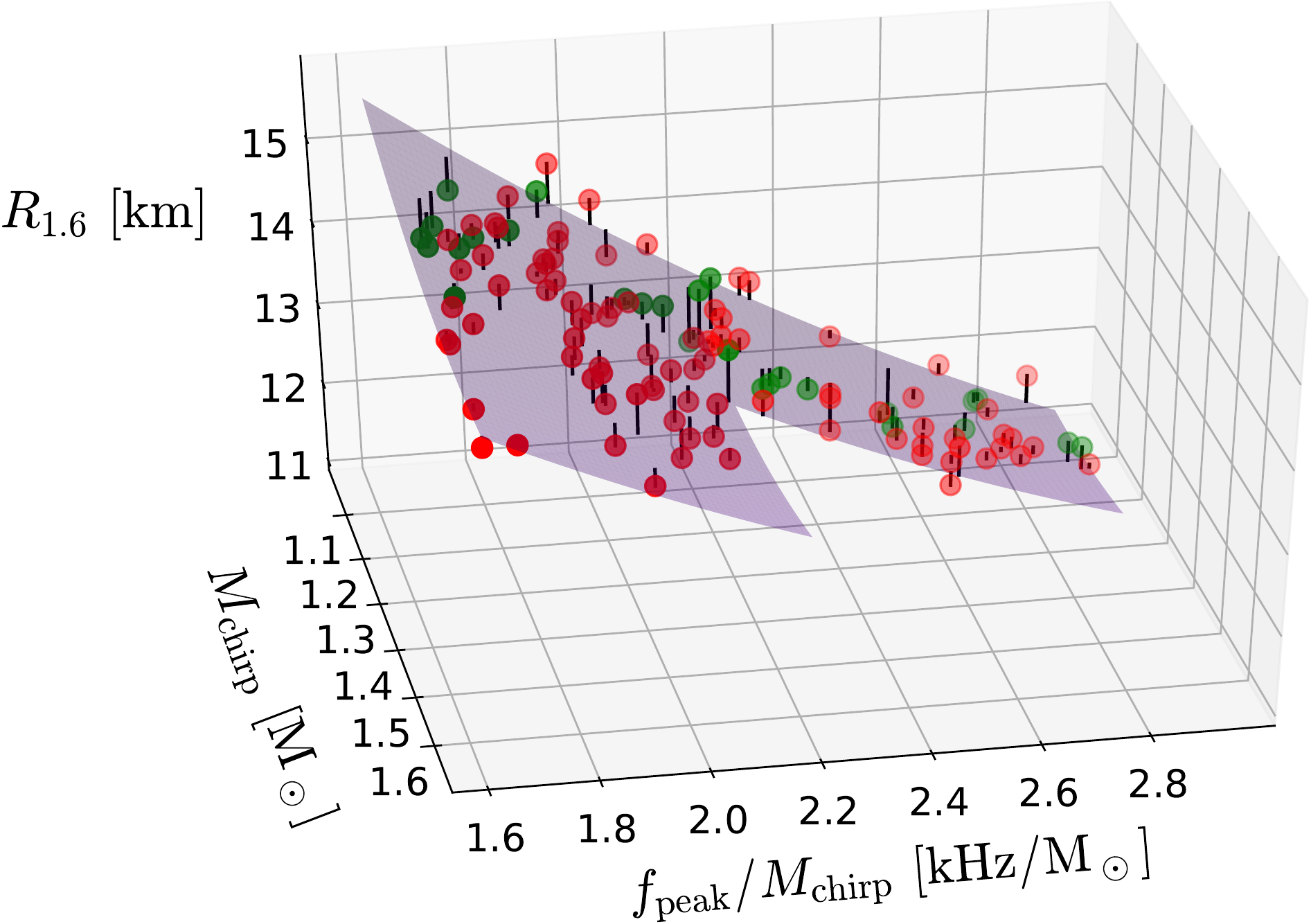}
  \caption{Empirical relation for $R_{1.6}$ using the whole set of models of the combined data set. Blue surface is the combined data sets surfaces. The red points correspond to Bauswein et. al. data and green points correspond to frequencies extracted from the CoRe GW catalogue.}
  \label{Combined_RfMsurfaces}
\end{figure}

Turning to the inverse empirical relations of the form $R_{\rm x}(f_{\rm peak}, M_{\rm chirp})$,  for $M = 1.6 M_\odot$ and using the subset of {\it equal-mass} configurations,
 the empirical relation for the radius
is\begin{equation}
  \begin{split}
    R_{1.6}= 39.258 -16.672 M_{\mathrm{chirp}} -10.784 f_{\mathrm{peak}}/M_{\mathrm{chirp}}\\
+3.952 M_{\mathrm{chirp}}^2 +2.75 f_{\mathrm{peak}} + 0.971 \left( f_{\mathrm{peak}}/M_{\mathrm{chirp}}
\right)^2,
  \end{split}
  \label{Combined_RfM_R16_e1}
\end{equation}
with a maximum residual of 0.605 km and $R^2=0.962$. 
 For the {\it whole set} of models (including both equal and unequal masses),
the empirical relation for the radius is
\begin{equation}
  \begin{split}
    R_{1.6}= 35.442 -13.46 M_{\mathrm{chirp}} -9.262 f_{\mathrm{peak}}/M_{\mathrm{chirp}}\\
+3.118 M_{\mathrm{chirp}}^2 +2.307 f_{\mathrm{peak}} +0.758 \left( f_{\mathrm{peak}}/M_{\mathrm{chirp}}
\right)^2,
  \end{split}
  \label{Combined_RfM_R16_e0}
\end{equation}
with a maximum residual of 0.654 km and $R^2=0.954$. The corresponding surface and data points are shown in  Fig. \ref{Combined_RfMsurfaces}. For $M=1.6M_\odot$ the
addition of the CoRe data points thus somewhat increases the maximum residual and this trend continues for lower masses, pointing to small systematic differences
due to  the different numerical treatments between the two data sets.

 Note that there are too few data points
for high-mass models in our chosen subset of CoRe models. We thus do not
construct a new relation for the radius of neutron stars with mass  $M=1.8M_\odot$.

\section{Empirical Relations for $f_{\rm peak}$ using tidal deformabilities}
\label{sec:fpeakL}

In \cite{CORE1}, an empirical
relation between  $f_{\rm peak} M_{\rm tot}$ and the dimensionless quadrupole tidal coupling constant 
\begin{equation}
\kappa_{2}^{\mathrm{T}} \equiv 2\left[\frac{1}{q}\left(\frac{X_{A}}{C_{A}}\right)^{5} k_{2}^{A}+q\left(\frac{X_{B}}{C_{B}}\right)^{5} k_{2}^{B}\right],
\end{equation}
was found, where $q:=M_A/M_B\geq 1$ is the mass ratio, $X_{A,B}:=M_{A,B}/M_{\rm tot}$, $k_2^{A,B}$ are the dimensionless quadrupole Love numbers and $C_{A,B}:=M_{A,B}/R$ is compactness {(see also \cite{Takami2015,Rezzolla2016})}.  Ref. \cite{2019PhRvD.100d4047T} reports that practically the same accuracy is achieved when using the mass-weighted tidal deformability  
\begin{equation}
\tilde{\Lambda}=\frac{16}{13} \frac{\left(M_{A}+12 M_{B}\right) M_{A}^{4} \Lambda_{A}+\left(M_{B}+12 M_{A}\right) M_{B}^{4} \Lambda_{B}}{\left(M_{A}+M_{B}\right)^{5}},
\end{equation}
in place of   $\kappa_2^T$  and an improvement is obtained by defining a new variable
\begin{equation}
\zeta:=\frac{3}{16}\tilde\Lambda+a \frac{M_{\rm tot}}{M^{\mathrm{TOV}}_{\rm max}},
\label{zetadef}
\end{equation} 
where  $a=-131.701$ (determined empirically by minimizing the RMS error) and $M^{\mathrm{TOV}}_{\rm max}$ is the maximum mass for nonrotating models allowed by a given EOS. The second term in (\ref{zetadef}) absorbs (to some degree) the mass dependence of the empirical relation found in \cite{CORE1} (see also \cite{Coughlin2018a}).  

The variable $\zeta$ used in the bivariate empirical relation in \cite{2019PhRvD.100d4047T} depends on the  tidal deformabilities of both stars, as well as on $M^{\mathrm{TOV}}_{\rm max}$. Determining $\zeta$ through a measurement of $f_{\rm peak}$ does not lead to a \textit{direct} constraint on the tidal deformability $\Lambda_{\rm x}$ at a specific mass (but indirect constraints could be inferred). A bivariate relation of the form $f_{\rm peak}(\Lambda_{\rm x})$ can be expected, since there exists a direct relation $\Lambda_{\rm x}(R_{\rm x})$, as demonstrated in \cite{2018PhRvL.120q2703A} for the particular case of $\Lambda_{\rm 1.4}$ (see also \cite{De2018}). Indeed, we find such a relation in Section \ref{LaX}.   

Even tighter empirical relations than the bivariate   $f_{\rm peak}(\Lambda_{\rm x})$ relation discussed in Section \ref{LaX} can be obtained by adding another variable, i.e.\ by constructing relations of the form  $f_{\rm peak}(\Lambda_{\rm x}, M_{\rm chirp})$. Such a multivariate relation can also be constructed using the mass-weighted tidal deformability $\tilde \Lambda$. We thus seek relations of the form   
\begin{equation}
  f_{\rm peak}M_{\rm chirp} = b_0 + b_1 M_{\rm chirp} + b_2 \Lambda^{-1/2},
  \label{eq:fLM:Ltilde}
\end{equation}
where $\Lambda$ is a placeholder for either $\tilde \Lambda$ or $\Lambda_{\rm x}$. The exponent of $-1/2$ in the last term was determined empirically.
We chose $M_{\rm chirp}$ instead of $M_{\rm tot}$ in \cite{CORE1,2019PhRvD.100d4047T}, since it is better constrained by observations. In this section we will only use the CFC/SPH dataset.

\subsection{Empirical relations using $\tilde{\Lambda}$}
\label{subsecempL}

\begin{figure*}[!]
\includegraphics[width=14cm]{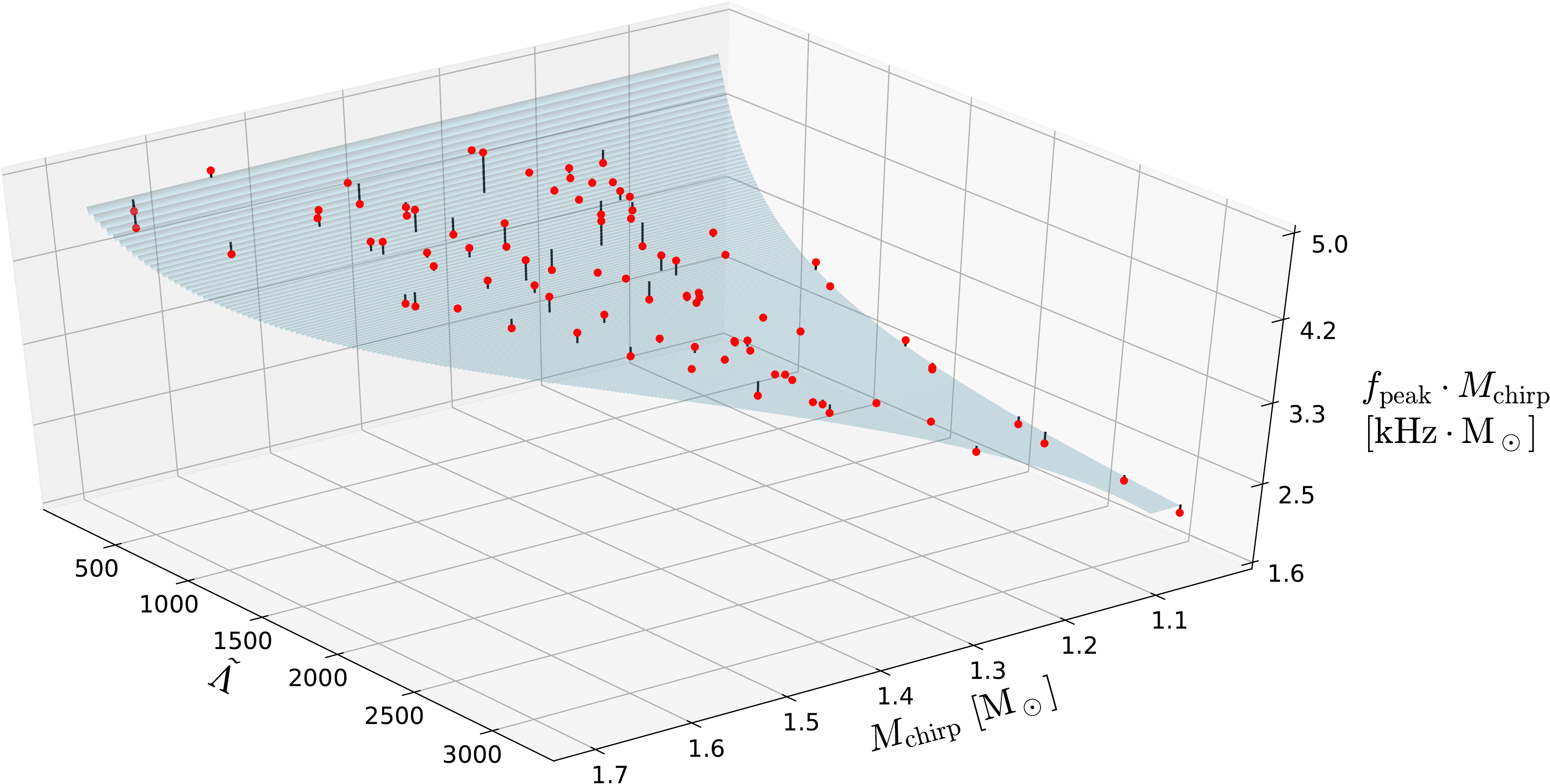}
 \caption{Empirical surfaces for $f_{\rm peak}$ using the chirp mass $M_{\rm
chirp}$ and the tidal deformability $\tilde{\Lambda}$. The red dots correspond
to the CFC/SPH data. The left figure corresponds to all models in the dataset
and the right figure corresponds to equal mass models only.}
 \label{fig:fLM:Ltilde}
\end{figure*}

For $\tilde{\Lambda}$ and using the  {\it whole set} of models, including
both equal and unequal mass configurations, the empirical relation for the
frequency $f_{\rm peak}$ is
\begin{equation}
    f_{\rm peak}M_{\rm chirp} = 1.392 - 0.108 M_{\rm chirp} + 51.70 \tilde{\Lambda}^{-1/2},
    \label{fpeak2DL}
\end{equation}
with a maximum residual corresponding to 0.302 kHz in terms of the frequency $f_{\rm peak}$
and $R^2=0.985$. The corresponding surface and data points are shown in 
Fig. \ref{fig:fLM:Ltilde}.
Restricting to equal-mass configurations, one obtains comparable (only slightly better) values for the maximum residual and $R^2$ of the fit. Moreover, restricting to  a bivariate relation of the type $f_{\rm peak}(\tilde \Lambda)$ (motivated by the bivariate relations found  in \cite{CORE1,2019PhRvD.100d4047T}) one obtains  a relation (the inverse of the $\tilde \Lambda(f_{\rm peak})$ fit discussed below in Section \ref{LtildeEmp}), which has a similar maximum residual and $R^2$ as for the multivariate fit (\ref{fpeak2DL}) and is comparable with the fits in \cite{CORE1,2019PhRvD.100d4047T}.
Thus, for the relation between $f_{\rm peak}$ and $\tilde \Lambda$ there exists no obvious advantage  in using a multivariate relation of the form (\ref{fpeak2DL}), but
this changes, when we consider tidal deformabilities at specific masses, $\Lambda_{\rm x}$, as we show below. 
 
Note that the accuracy can increase significantly, if one considers setups with a fixed total binary mass. In \cite{Bauswein2019a} the maximum residual when using
$\Lambda_{1.35}$ was found to be only of order 100Hz for symmetric binaries of 1.35+1.35$M_\odot$ employing a large set of purely hadronic EOS.   

\subsection{Empirical relations using  different $\Lambda_{\rm x}$}

We construct empirical relations for $f_{\rm peak}$ using different $\Lambda_x$, where $x=1.4, 1.6$ and 1.8. Here, we present the relations only for the whole set of models, including both equal and unequal-mass models (restricting to equal-mass models only, yields slightly better fits). 

For $\Lambda_{1.4}$ 
the empirical relation is 
 \begin{equation}
    f_{\rm peak}M_{\rm chirp} = -4.015 + 4.490 M_{\rm chirp} + 47.14 \Lambda_{1.4}^{-1/2},
     \label{fpeak2DL1.4}
\end{equation}
with a maximum residual of 0.452
kHz in terms of the frequency $f_{\rm peak}$
and $R^2=0.971$.
We note that neglecting the exponent of $-1/2$ in the last term of  (\ref{fpeak2DL1.4}) gave a slightly better fit, but we keep this exponent for uniformity with the corresponding relations for higher masses.

\begin{figure*}[!]
 \includegraphics[width=8.5cm]{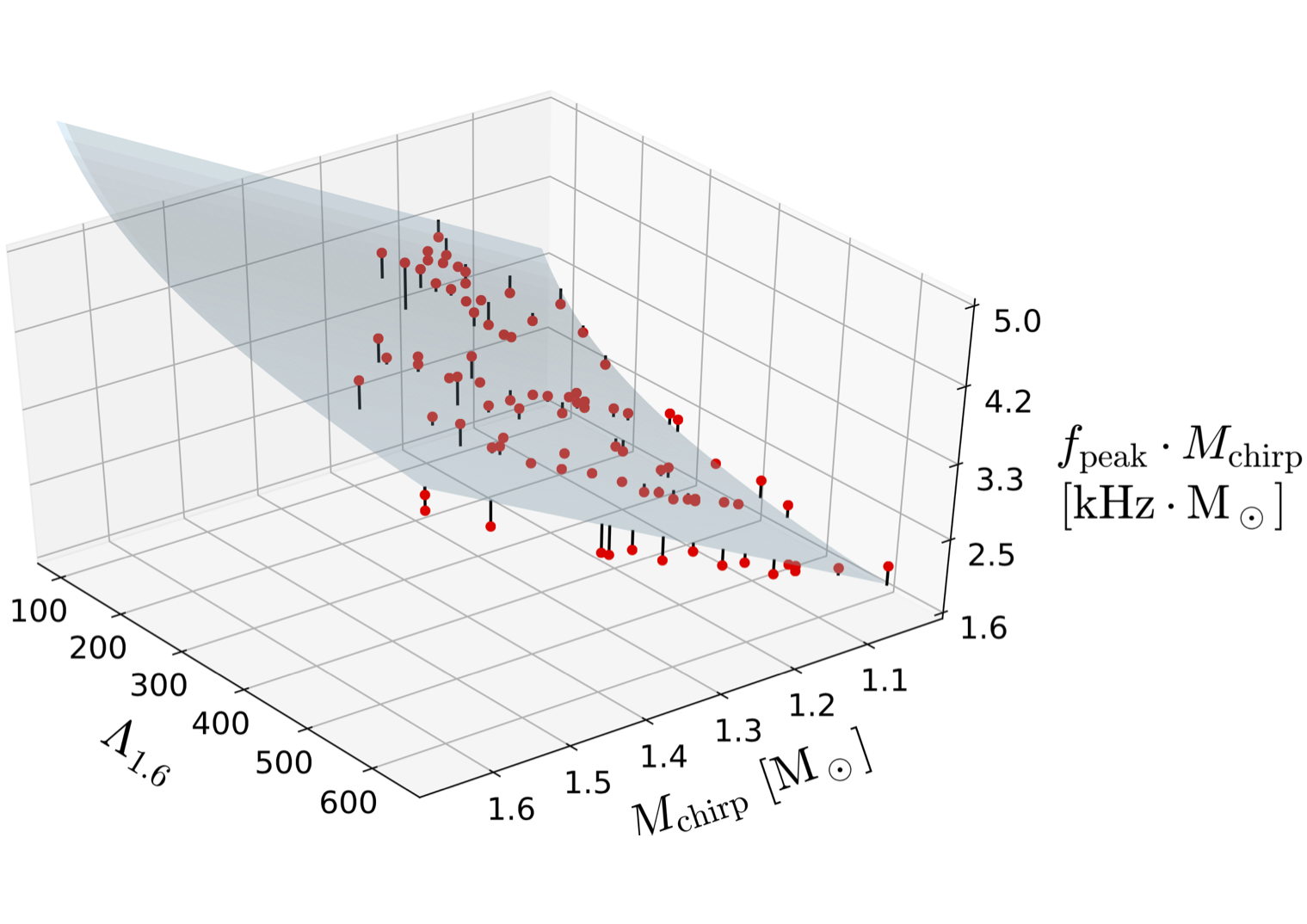}
 \includegraphics[width=8.5cm]{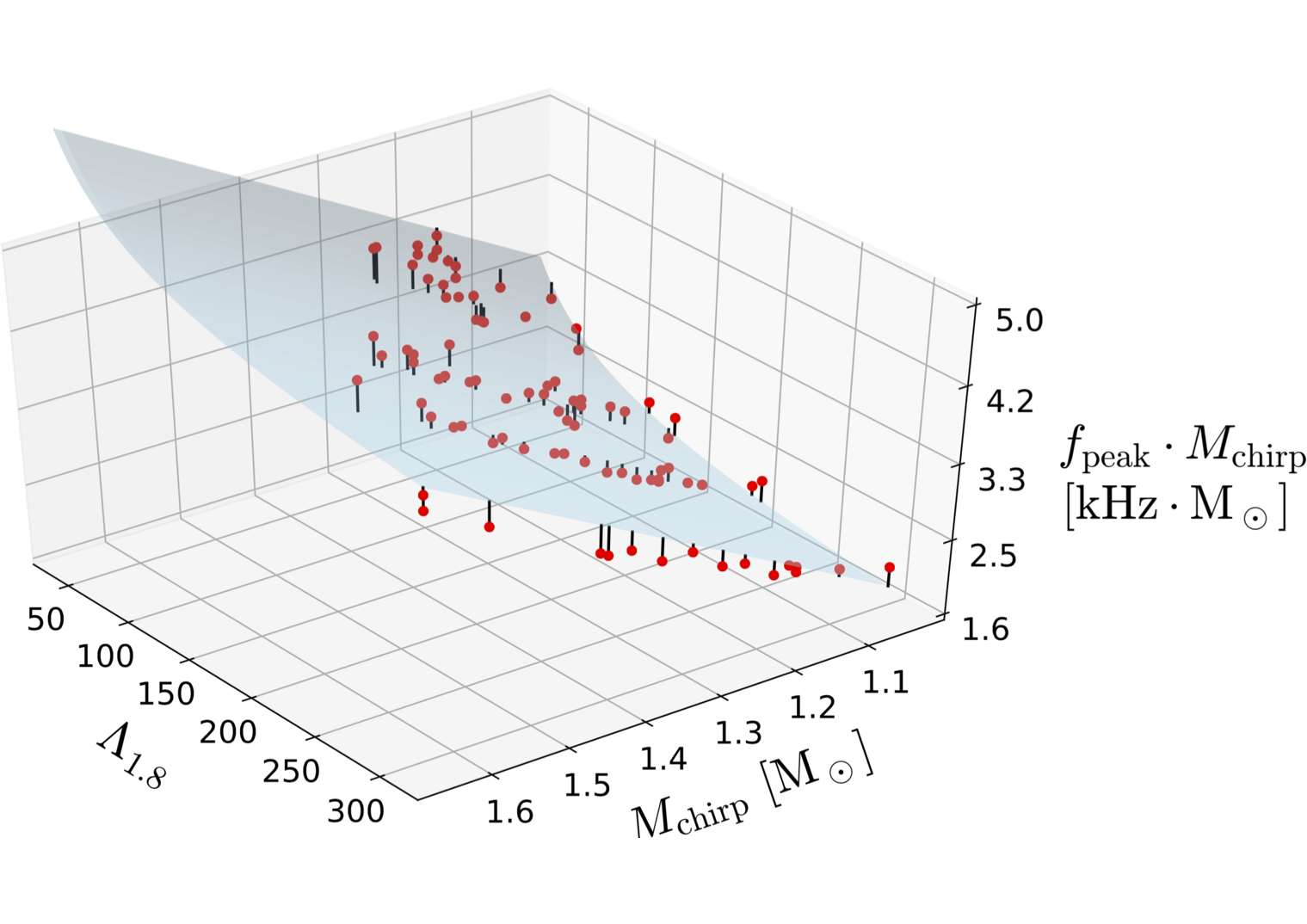}
 \caption{{\it Left panel:} Multivariate empirical relation for $f_{\rm peak}$ using the tidal deformability $\Lambda_{1.6}$ and the chirp mass $M_{\rm chirp}$. The red dots correspond to the CFC/SPH data. {\it Right panel:} Same as left panel, but with $\Lambda_{1.8}$}
 \label{fig:fLM:L16}
\end{figure*}

For $\Lambda_{1.6}$ the empirical relation is  \begin{equation}
    f_{\rm peak}M_{\rm chirp} =  -3.922 + 4.528 M_{\rm chirp} +28.35  \Lambda_{1.6}^{-1/2},
\end{equation}
with a maximum residual of 0.373 kHz in terms of the frequency $f_{\rm peak}$
and $R^2=0.973$ (see left panel of Fig. \ref{fig:fLM:L16}) and for $\Lambda_{1.8}$ the empirical relation is
\begin{equation}
    f_{\rm peak}M_{\rm chirp} =  -3.73 + 4.548 M_{\rm chirp} +15.94  \Lambda_{1.8}^{-1/2},
\end{equation}
with a maximum residual of 0.283 kHz in terms of the frequency $f_{\rm peak}$
and $R^2=0.967$ (see right panel of Fig. \ref{fig:fLM:L16}).

\section{Empirical relations for Tidal Deformabilities using $f_{\rm peak}$}
\label{sec:Lfpeak}

We construct
multivariate and empirical relations for the tidal deformabilities  $\tilde{\Lambda}$ and $\Lambda_{x}$ with $x=1.4,1.6$ and 1.8. The relation for $\tilde{\Lambda}$ has
the form of
\begin{equation}
  \tilde{\Lambda} = b_0 + b_1 M_{\rm chirp} f_{\rm peak} + b_2 f_{\rm peak}^{-2},
  \label{eq:LfM:Ltilde}
\end{equation} 
whereas the relations for different $\Lambda_{x}$ are of the form
\begin{equation}
  \Lambda_{\rm x} = b_0 + b_1 M_{\rm chirp} + b_2 f_{\rm peak} + b_3 f_{\rm
peak}^2, 
  \label{eq:LfM:Lx}
\end{equation}
(the above forms represent optimal choices among a number of different versions that we investigated).

In addition, we explore bivariate relations of the form $\tilde \Lambda (f_{\rm peak}M_{\rm chirp})$ and $ \Lambda_{x}
(f_{\rm peak}/M_{\rm chirp})$, in which the product $f_{\rm peak}M_{\rm chirp}$ or the ratio  $f_{\rm peak}/M_{\rm
chirp}$, correspondingly, are treated as  a single variable. 

\subsection{Empirical relations for $\tilde{\Lambda}$}
\label{LtildeEmp}

\begin{figure*}
 \includegraphics[width=16cm]{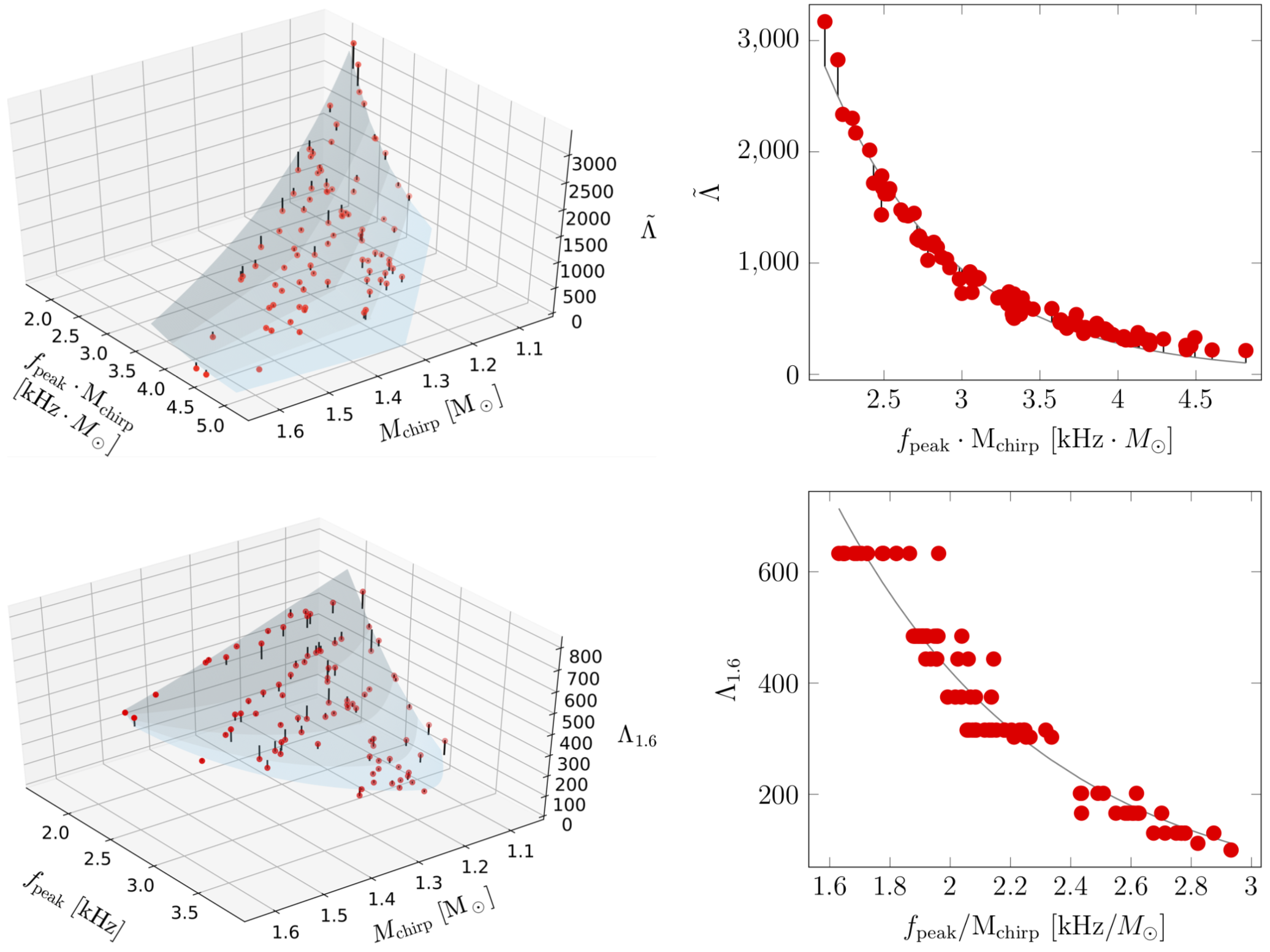}
\caption{\textit{Top row:} Multivariate empirical relations (left)\ and bivariate empirical relations (right) for $\tilde{\Lambda}$. Both have comparable accuracy. \textit{Bottow row:} Multivariate (left)\ and bivariate (right) empirical relations
for $\Lambda_{1.6}$. The multivariate relation has a significantly smaller maximum residual than the bivariate relation. Red dots correspond to the
CFC/SPH data.  }
 \label{fig:LfM:Ltilde2D}
\end{figure*}

For $\tilde{\Lambda}$ and using the subset of {\it equal-mass} configurations,
the empirical relation using the frequency $f_{\rm peak}$ is
\begin{equation}
 \tilde{\Lambda} = -1434 + 120.1 M_{\rm chirp} f_{\rm peak} +  18053
f_{\rm peak}^{-2},
\end{equation}
with a maximum residual of 315.8  and
$R^2=0.985$, whereas using the the {\it whole set} of models, including
both equal and unequal mass configurations, the empirical relation is
\begin{equation}
    \tilde{\Lambda} =  -1344 + 108.9 M_{\rm chirp} f_{\rm peak} + 17208
 f_{\rm peak}^{-2},
\end{equation}
with a maximum residual of 433.1 and $R^2=0.975$
(see  top left panel of Fig. \ref{fig:LfM:Ltilde2D}).

For $\tilde \Lambda$ we also construct a bivariate empirical relation of the form
\begin{equation}
\tilde{\Lambda} = b_0 e^{-z/b_1}, 
\label{tildeLuni}
\end{equation}
where the product $z=f_{\rm peak} M_{\rm chirp}$ is treated as a single variable (this is motivated by the existence of bivariate relations of the form $z(
\kappa_2^T)$ in \cite{CORE1} and  $z(\tilde \Lambda)$ or $z(\zeta)$ in \cite{2019PhRvD.100d4047T}, but we use a different functional form of the fit, that gave a smaller residual). 

For the subset of {\it equal-mass} configurations,
we find $b_0=0.836$ and $b_1=36014$,
with a maximum residual of 325.5 and $R^2=0.979$, whereas for 
the {\it whole set} of models, including
both equal and unequal mass configurations, we find $b_0=0.817$ and $b_1=37096$,
with a maximum residual of 403.1 and $R^2=0.969$ (see  top right panel of Fig. \ref{fig:LfM:Ltilde2D}).

We thus find that the bivariate empirical relation of the form (\ref{tildeLuni}) is of comparable accuracy as the multivariate empirical relation of the form (\ref{eq:LfM:Ltilde}) and the latter does not have an advantage over the former, as anticipated by the results of Section
\ref{subsecempL}.
\subsection{Empirical relations for $\Lambda_{x}$}
\label{LaX}

Next, we construct multivariate empirical relations of the form (\ref{eq:LfM:Lx}) for  different $\Lambda_x$,
where $x=1.4, 1.6$ and 1.8. Here, we present the relations only for the whole
set of models, including both equal and unequal-mass models (restricting
to equal-mass models only yields fits of comparable accuracy). 

The empirical relations we construct are:\begin{equation}
  \Lambda_{1.4} = 5083+ 1588M_{\rm chirp} - 3787 f_{\rm peak}
+ 535.7 f_{\rm peak}^2, 
\label{L1.4}
\end{equation}
(maximum residual of 185.4 and $R^2=0.958)$,
\begin{equation}
   \Lambda_{1.6} = 2417 + 770.2 M_{\rm chirp} - 1841 f_{\rm peak}
+ 262.9 f_{\rm peak}^2,
\label{L1.6}
\end{equation}
(maximum residual of 99.85 and $R^2=0.964$,
see bottom left panel of Fig. \ref{fig:LfM:Ltilde2D}),
and\begin{equation}
   \Lambda_{1.8} = 1253 + 398.7 M_{\rm chirp} - 982.8 f_{\rm peak}
+ 143.2 f_{\rm peak}^2,
\label{L1.8}
\end{equation}
(maximum residual of 74.35 and $R^2=0.933)$.
Notice that the maximum residual in $\Lambda_x$ is getting smaller as the target mass increases.

Finally, we construct bivariate empirical relations of the form $\Lambda_x(u)$, where 
the ratio $u=f_{\rm peak}/ M_{\rm chirp}$ is treated as a single variable. The empirical relations are
\begin{equation}
  \Lambda_{1.4} = 12845 e^{-u/0.77},
  \label{bL1.4}
\end{equation}
(maximum residual of 345.4 and $R^2=0.92)$,
\begin{equation}
   \Lambda_{1.6} =  7251e ^{-u/0.703},
   \label{bL1.6}
\end{equation}
(maximum residual of 187.4 and $R^2=0.931$, see bottom right panel of Fig. \ref{fig:LfM:Ltilde2D}), and
\begin{equation}
   \Lambda_{1.8} = 4977e^{-u/0.612},
   \label{bL1.8}
\end{equation}
(maximum residual of 107.5 and $R^2=0.911)$.

The multivariate empirical fits (\ref{L1.4}) $-$ (\ref{L1.8}) have a maximum residual for $\Lambda_x$ that is consistently roughly half of the corresponding maximum residual for the bivariate fits (\ref{bL1.4}) $-$ (\ref{bL1.8}) This allows for an accurate determination of the tidal deformability at specific masses, $\Lambda_x$, through the observables $f_{\rm peak}$ and $M_{\rm chirp}$, which would then place direct constraints on the EOS. This is complementary (and of similar accuracy)  to the accurate determination of radii at specific masses, $R_x$, which we presented in Sections \ref{sec:rad} and \ref{sec:mr}.

We note that further reduction of the maximum residual can be attained for certain fixed chirp masses (or fixed total masses), essentially taking slices of the \ empirical surface in Fig. \ref{fig:LfM:Ltilde2D} for fixed $M_{\rm chirp}$ (which will be known to high accuracy from the inspiral phase). Such relations for fixed binary setups as
shown in [39] should be ultimately used for constraints on the tidal
deformability from $f_\mathrm{peak}$ because they yield the smallest
scatter, which determines the systematic error. Binary masses can be
accurately measured for events where postmerger GWs are detectable.

\section{Discussion and conclusions}\label{sec:sum}

In this paper we explore empirical relations for distinct postmerger GW frequencies of BNSs such that they can be directly implemented in GW data analysis procedures for parameter estimation. These frequencies are extracted from a large representative sample of BNS merger simulations for different binary mass configurations and model EOSs. We employ results from two different catalogues of simulations, which are based on different numerical codes. We focus on relations between postmerger GW frequencies, the chirp mass of the binary system and NS radii. The latter are determined by the incompletely known EOS, and we investigate radii of different fiducial NS masses to characterize different density ranges of the EOS.

Since the binary mass ratio $q$ may not be measured with high precision, our complete set of models includes binaries within a relatively large range of mass ratios.  To approximately assess the impact of the mass ratio, we derive empirical relations also for equal-mass mergers only and find unsurprisingly tighter relations. This demonstrates that, if available, information on $q$ should be included in such empirical relations.

Because we aim at GW data analysis applications, we derive two separate sets of relations. Once, the GW frequencies are dependent variables. This type of relations can be implemented to predict the expected postmerger GW signal for given EOS models (Sect.~\ref{sec:freq}) and may be linked to EOS information from the GW inspiral phase. The maximum residuals found for our relations may be used to quantify the uncertainties (or to define priors in other types of analyses). For another set of relations, NS radii are treated as dependent variables. These relations can be employed to determine NS radii from the measurement of postmerger GW frequencies (Sect.~\ref{sec:rad}).

By using a large sample of BNS simulations we can assess the quality of the individual empirical relations, which we obtain by least-square fits. We quantify the accuracy of these relations by the maximum residual. This deserves a comment. The maximum residual is the most meaningful figure of merit for an empirical relation because any other statistical measure could be strongly biased by the chosen sample of underlying models. This is because the data for constructing the fits do not follow a statistical distribution, but they are simply given by the available models for the EOS and chosen simulation setups. We caution that even if one uses some sort of parametrization of the EOS, it is not obvious that one can employ other statistical measures to assess the quality of an empirical relation. It is not clear which distribution the parameters should follow in order to be representative unless they can be physically motivated. Moreover, the space of EOS parameters is mapped in a non-trivial way on NS properties and GW frequencies. Obviously, also the maximum residual depends on the underlying data. But we expect that by employing a very large sample of models, the data will contain the most extreme outliers. Then, the maximum residual provides a meaningful upper limit on the uncertainties and by how much the true value could at most deviate from the fit.  

Our main findings can be summarized as follows.

(1) We find generally tight relations between postmerger GW frequencies, the chirp mass and NS radii. Typically maximum residuals are of the order of 300~Hz (or a few hundred meters if NS radii are the dependent quantity).

(2) Apart from tight relations for the dominant postmerger GW frequency, we confirm the existence of two separate  empirical relations for two distinct subdominant peaks of the postmerger GW spectrum, in agreement with~\cite{Bauswein2015,Bauswein2016}. These findings are in tension with the interpretation of~\cite{Takami2014,Takami2015} that a single universal function is sufficient to describe the behavior of subdominant peaks in the postmerger GW spectrum. (Slight disagreements of up to a few 100~Hz between the frequencies of secondary peaks predicted by fit formulae in~\cite{Bauswein2015} on one hand and the data in~\cite{Rezzolla2016} on the other hand are fully compatible with the scatter of the fit formulae in~\cite{Bauswein2015} and the maximum residuals we observe in this study for a larger set of models.) 
The existence of two distinct subdominant peaks, and thus corresponding relations, is impressively corroborated by a machine learning algorithm, which  identified three different classes of postmerger spectra in remarkable agreement with the classification scheme introduced in~\cite{Bauswein2015}.

The machine-learning method employed here may be used for an automated identification of the type of postmerger spectrum in numerical simulations or in future GW data analysis application.

(3) For most relations investigated here those with the dominant postmerger frequency $f_\mathrm{peak}$ yield the smallest maximum residual in comparison to relations where the subdominant peaks were used. This stresses the importance of $f_\mathrm{peak}$ for EOS constraints, considering also that secondary peaks may be harder to measure (because of their lower signal-to-noise ratio) and may yield larger statistical errors in a measurement because of their generally larger width in comparison to $f_\mathrm{peak}$.

(4) Our study also confirms that radii of high-mass NSs are more suitable to describe the EOS dependence of postmerger frequencies~\cite{Bauswein2012a}. We compare empirical relations for $R_{1.2}$, $R_{1.4}$, $R_{1.6}$ and $R_{1.8}$, i.e.\ we characterize a given EOS by the radius of nonrotating NSs with different masses of 1.2~$M_\odot$, 1.4~$M_\odot$, 1.6~$M_\odot$ and 1.8~$M_\odot$. The radii of nonrotating NSs with different masses represent integral characteristics of the EOS in different density regimes, i.e.\ high-mass NSs reflect the EOS behavior at higher densities. Empirical relations for postmerger frequencies with $R_{1.6}$ or $R_{1.8}$ lead to systematically smaller maximum residuals considering the full range of binary masses. This behavior had already been observed in~\cite{Bauswein2012a} and explained by the fact that during merging the densities increase, which is why high-mass NSs better represent the density regime of the postmerger remnant and thus its GW emission. The confirmation of this finding is important because the inspiral GW signal of a BNS constrains the EOS regime of the two coalescing stars  stars, which in most cases are expected to be NSs with moderate masses. Moreover, the finite-size effects of high-mass NSs decrease in magnitude and are thus harder to measure with good accuracy. Hence, measuring radii of high-mass NSs through the postmerger phase provides complementary information on the high-density regime of the EOS.

 (5) Constructing different fits in this work, we recognize, not unexpectedly, that it is meaningful to restrict the parameter range, because this leads to tighter relations and thus implies smaller uncertainties in applications of these relations, e.g.\ for radius measurements. For instance, we find that considering only equal-mass mergers leads to tighter fits. In this context we briefly comment on the analysis of~\cite{Kiuchi2019a} who  find a somewhat larger scatter between $f_\mathrm{peak}$ and NS radii compared to previous results. This observation, however, is entirely a consequence of including unequal-mass binary configurations as well as equal-mass binaries with a variation of the binary mass ratio comparable to that inferred for GW170817 (i.e.\ between about 0.7 and 1.0). Considering for instance only the equal-mass results of~\cite{Kiuchi2019a} yields similarly tight relations as in previous studies. In future events which will allow the extraction of postmerger GW frequencies, the binary mass ratio can be expected to be measured with significantly better precision than for GW170817~\cite{Farr2016}.  We thus expect that empirical relations of the form we employ here, specialized to certain ranges for the binary mass ratio, would still have maximum residuals comparable to what is found in previous and the present works. 

We also emphasize that Ref.~\cite{Kiuchi2019a} use a simplified description of the NS crust. The crust EOS in fact is known with better precision.  This explains a quantitative bias between the GW frequencies in~\cite{Kiuchi2019a} compared to the previous fitting formulas in~\cite{Bauswein2012a} which are based on a proper description of the crust material. We expect that GW frequencies are, to good approximation, unaffected by the description of the crust EOS, while TOV solutions and thus NS radii do change by several 100m if a simplified NS crust is employed. Correcting this systematic underestimation of NS radii by the crust treatment, one finds that the equal-mass data of~\cite{Kiuchi2019a} are in excellent agreement with the fit formula in~\cite{Bauswein2012a}. We remark that a similar issue arises for the relations presented in~\cite{Takami2014,Takami2015,Rezzolla2016}, where the simplified crust treatment also introduces a quantitative bias, which implies that the resulting fit formulae cannot be directly applied for comparisons or for radius/EOS constraints. Instead the systematic shift of the TOV solutions should be removed for real applications.


Restricting our sample of models to a smaller range in the chirp mass, yields smaller maximum residuals. This is not unexpected considering previous results in the literature, which often focused on fixed binary mass configurations and found generally smaller deviations. Recall that the chirp mass is measured with very good precision from the GW inspiral phase.

We also anticipate that including additional constraints on the possible EOSs will result in more accurate fits with smaller maximum residuals. We do not further elaborate these considerations because in this study we want to quantify the maximum possible deviations from empirical relations for the postmerger GW emission. We expect to obtain robust upper limits by considering the largest possible set of models, which likely includes the most extreme, and possibly unrealistic, cases. We thus study here the worst-case scenario and stress that in future measurements significant improvements are anticipated if additional limits on the parameter range (mass ratio, chirp mass, EOS) are taken into account. 

Notice that a few of the EOS we use are somewhat (but not dramatically)\ disfavoured by the inferred
EOS constraints by the GW170817 event (the radii for typical neutron star
masses are about 1km larger than the 90\% credibility constraints in \cite{PhysRevLett.121.161101}).
When tighter EOS constraints from the inspiral phase (or from other observational methods) become available, then this will reduce the available parameter space, leading to improved empirical relation for the post-merger phase.


(6) As another important step to assess the maximum residuals and the quality and reliability of empirical relations for the dominant postmerger frequency $f_{\rm peak}$,  we construct fits based on two independent catalogues of models (CFC/SPH and CoRe). We do not find significant systematic differences between the two data sets, which is important because the codes are based on different numerical methods and slightly differ with regard to the implemented physical model. We also observe that the maximum residuals do not appreciably change if we include the second data set to our baseline models (CFC/SPH). This may indicate that the maximum residuals determined in this study are approximately converged.

(7) We confirm the existence of a bivariate empirical relation for $\tilde \Lambda$, e.g.~\cite{CORE1,Takami2015,2019PhRvD.100d4047T}. For the tidal deformability at specific masses $\Lambda_x$ (which is related to the radius at specific masses $R_x$) we find accurate \textit{multivariate} empirical relations, which can lead to tight constraints on the EOS.
The
empirical
relations involving the tidal deformability can actually be improved by fixing the chirp mass (or total binary mass), as demonstrated in \cite{Bauswein2019a}.

We conclude by mentioning a few caveats of our study and describe directions of future research. The data sets which we employ for constructing fits are based on a large sample of models but not on a systematic variation of the model parameters, which is in particular for the EOS not trivial to realize. Hence, the derived fits as well as the corresponding maximum residuals may be to some extent biased by the available models for instance because 1.35-1.35~$M_\odot$ binary models are over-represented, as a very common configuration. It may be interesting to choose merger models in a more systematic way, to check whether the current study is prone to selection effects. We also emphasize that the occurrence of a strong first-order phase transition (no included in the present study) can lead to a significant increase of the postmerger frequencies and thus to  deviations from the empirical relations which are based on models without strong phase transitions \cite{Bauswein2019a}. This also deserves more attention in future work. Finally, this study highlights the potential of machine learning for the recognition of specific types of postmerger spectra, which are linked to the underlying dynamics. Future work should explore whether these algorithms work in GW data analysis of actual events.

\acknowledgements
We are thankful to Luca Baiotti,\ Gabriele Bozzola, Tim Dietrich, Jocelyn Read and Kostas Kokkotas for comments. AB acknowledges support by the European Research Council (ERC) under the European Union’s Horizon 2020 research and innovation programme under grant agreement No. 759253, by the Sonderforschungsbereich SFB 881 ``The Milky Way System'' and the Sonderforschungsbereich SFB 1245 ``Nuclei: From fundamental interactions to structure and stars'' of the German Research Foundation (DFG) and the Klaus-Tschira Foundation. NS is supported by the ARIS facility of GRNET in Athens (GWAVES, GRAVASYM and SIMGRAV allocations). We are grateful for networking support through the COST\ actions CA16214 “PHAROS” and CA16104 “GWVerse”, CA17137  “G2Net” and CA18108 “QG-MM”.
\bibliography{VSB}

\appendix
\newpage

\section{CFC/SPH GW catalogue.}
\label{Appendix.A}
\begin{figure}[H]
\centering
 \includegraphics[width=8.5cm]{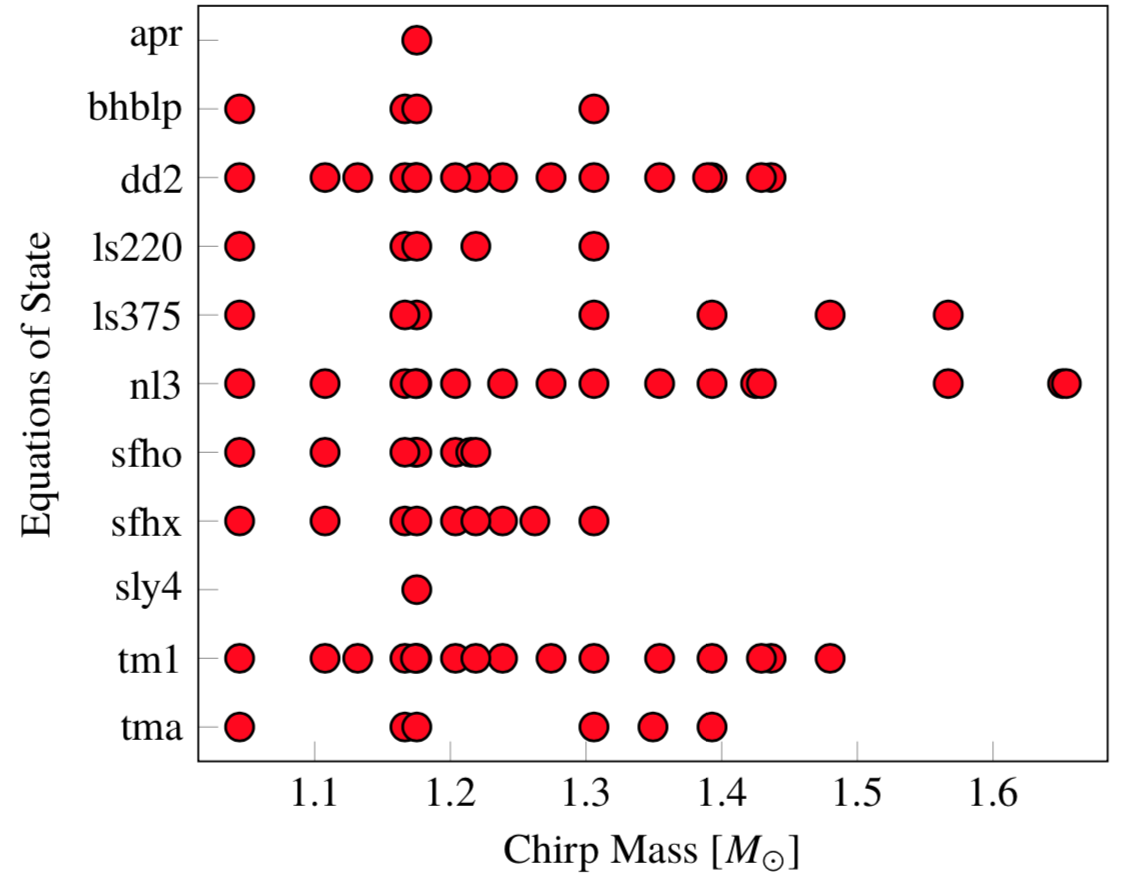}
\caption{EOS vs. chirp mass for the models comprising the  CFC/SPH GW catalogue.} \label{fig:eoschirp}
\end{figure}

\begin{figure}[H]
\centering
 \includegraphics[width=8.5cm]{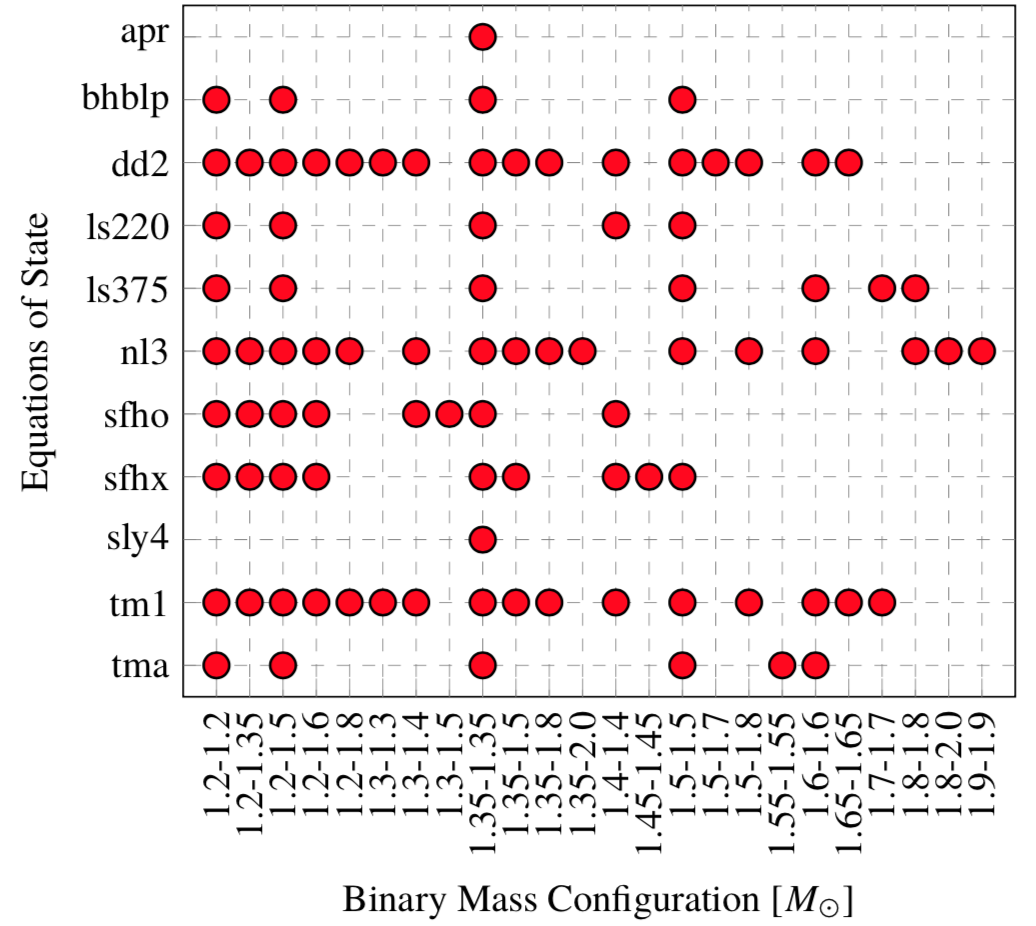}
\caption{EOS vs. binary mass configuration for the models comprising the  CFC/SPH GW catalogue.} \label{fig:eosconf}
\end{figure}

The models comprising the  CFC/SPH GW catalogue are shown in a grid of EOS vs. chirp mass (Fig. \ref{fig:eoschirp}) and of EOS vs. individual masses of the binary system  (Fig. \ref{fig:eosconf}). 

\section{CoRe catalogue}
\label{Appendix.B}
\begin{figure}[H]
 \includegraphics[width=8.5cm]{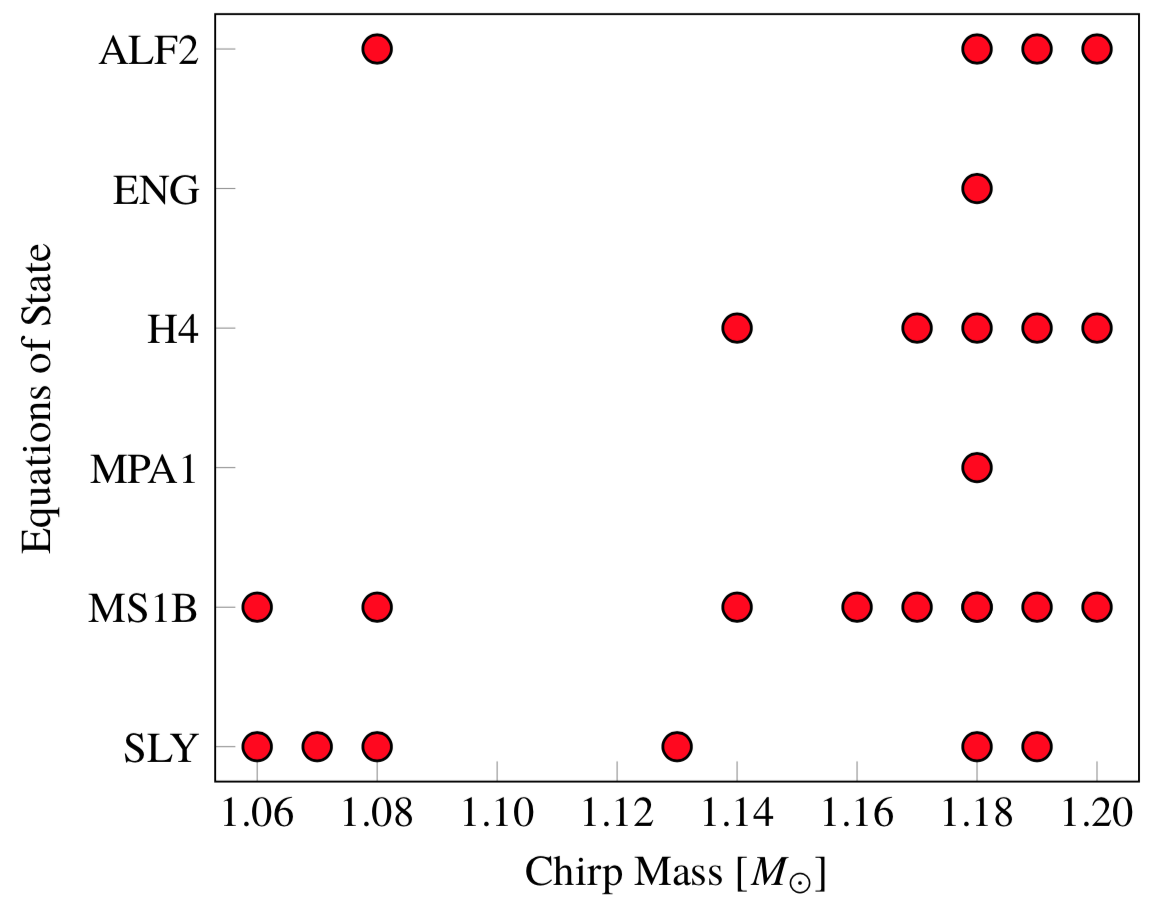}
 \caption{EOS vs. chirp mass for the subset of models from the CoRe GW catalogue used in the present study.}
 \label{fig:COREeoschirp}
\end{figure}

\begin{figure}[H]
 \includegraphics[width=8.5cm]{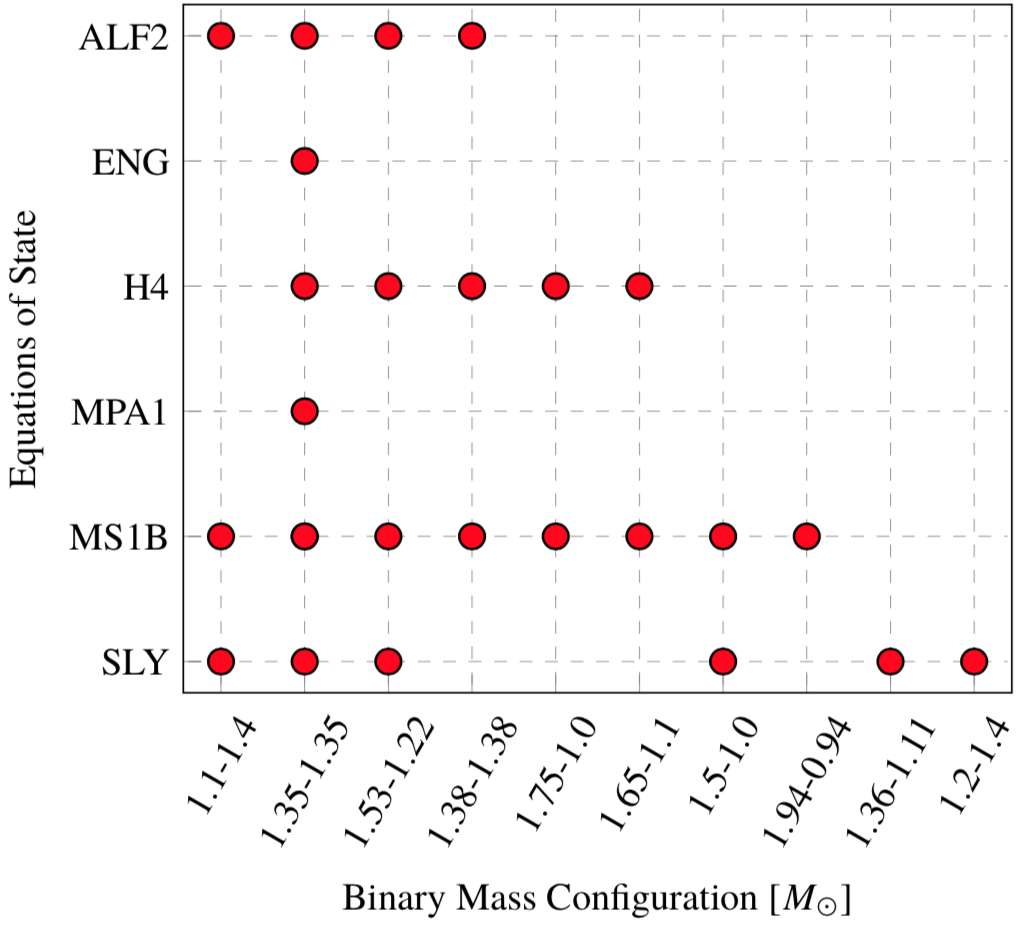}
\caption{EOS vs. mass configuration of the binary for the subset of models from the CoRe GW catalogue
used in the present study.}
\label{fig:COREmodels}
\end{figure}

The models comprising the  subset of the CoRe GW catalogue used in the present study are shown in a grid of EOS
vs. chirp mass (Fig. \ref{fig:COREeoschirp}) and of EOS vs. individual masses
of the binary system  (Fig. \ref{fig:COREmodels}). 

\section{Equilibrium properties of nonrotating models}
\label{Appendix.C}
Table \ref{table:TOV-info} summarizes equilibrium properties of nonrotating models for all EOS used in the present study. Specifically, the radius of isolated, cold equilibrium models of mass $M=1.2, 1.4, 1.6$ and 1.8$M_\odot$\ is shown. 
\begin{table}[H]
  \centering
\caption{Radius $R_{\rm x}$ for specific masses (shown as subscript in units of solar mass)  for the EOS used in the present study.}
  \begin{tabular}{|c|c|c|c|c|}
    \hline
EoS & ${R}_{1.2}$ &${R}_{1.4}$ & ${R}_{1.6}$ & ${R}_{1.8}$ \\  \hline
APR          & 11.37           & 11.34           & 11.27           & 11.14           \\ \hline
BHBLP        & 13.08           & 13.14           & 13.12           & 12.95           \\ \hline
DD2          & 13.08           & 13.15           & 13.18           & 13.16           \\ \hline
LS220        & 12.65           & 12.56           & 12.36           & 12.00           \\ \hline
LS375        & 13.43           & 13.57           & 13.67           & 13.72           \\ \hline
NL3          & 14.64           & 14.69           & 14.73           & 14.75           \\ \hline
SFHO         & 11.89           & 11.81           & 11.68           & 11.43           \\ \hline
SFHX         & 11.87           & 11.92           & 11.90           & 11.79           \\ \hline
TM1          & 14.44           & 14.38           & 14.26           & 14.06           \\ \hline
TMA          & 13.75           & 13.72           & 13.57           & 13.24           \\ \hline
ALF2         & 12.64           & 12.70           & 12.64           & 12.40           \\ \hline
ENG          & 11.97           & 11.97           & 11.92           & 11.79           \\ \hline
H4           & 14.04           & 13.95           & 13.74           & 13.32           \\ \hline
MPA1         & 12.37           & 12.43           & 12.47           & 12.46           \\ \hline
MS1b         & 14.43           & 14.53           & 14.60           & 14.65           \\ \hline
SLy          & 11.80           & 11.71           & 11.55           & 11.27           \\ \hline
\end{tabular}
\label{table:TOV-info}
\end{table}

\section{Frequencies}
\label{Appendix.D}
Table \ref{table:cfc/sph-data} lists the extracted postmerger oscillation frequencies ($f_{\rm peak}$, $f_{2-0}$, $f_{\rm spiral}$) along with the individual masses $m_1$, $m_2$, the chirp mass $M_{\rm chirp}$ and the mass ratio $q$ for the models comprising the CFC/SPH GW catalogue. Table \ref{table:CORE-data} lists $f_{\rm peak}$, $m_1$, $m_2$, $M_{\rm chirp}$ and $q$ for the models comprising the  subset of the CoRe GW catalogue used in the present study (in addition, $f_{\rm gw}$ is the GW frequency of the binary system at the start of the simulation, as defined in the CoRe GW catalogue).

\section{Regression information for the empirical relations for frequencies and radii constructed with the CFC/SPH GW catalogue}
\label{Appendix.E}
Table \ref{table:fRM} lists the detailed information for the empirical relations of the form (\ref{fRM}) $f_j(R_{\rm x}, M_{\rm
chirp})$ for the three different postmerger frequencies $f_{\rm peak}$,  $f_{2-0}$ and $f_{\rm spiral}$.  $R_{\rm x}$ is the radius of nonrotating models of mass $x$ (in $M_\odot$). The table includes the  values of coefficients $b_0,..., b_5$ of the fit, the adjusted $R^2$ of the fit, the maximum and the mean residual as well as the standard deviation of the residuals. 

Table \ref{table:RfM}  lists the detailed information for the empirical relations of the form (\ref{RfM})
$R_{\rm x}(f_j, M_{\rm chirp})$. The parameter $f_j$ stands for the three different postmerger frequencies $f_{\rm peak}$, 
$f_{2-0}$ and $f_{\rm spiral}$. In all cases, we show fits for either the subset of models where the two components of the binary system have equal masses or for the whole set of models. Note that $R_{\rm x}^<$  ($R_{\rm x}^>$) stands for fits where only the low-mass (high-mass) portion of the data was used. The table includes the
 values of coefficients $b_0,..., b_5$ of the fit, the adjusted $R^2$ of
the fit, the maximum and the mean residual as well as the standard deviation
of the residuals.
\newpage
\begin{table*}[h] 
  \centering
  \caption{Extracted postmerger oscillation
frequencies ($f_{\rm peak}$, $f_{2-0}$, $f_{\rm spiral}$) along with the
individual masses $m_1$, $m_2$, the chirp mass $M_{\rm chirp}$ and the mass
ratio $m_1/m_2 \leq 1$ for the models comprising the CFC/SPH GW catalogue.}
  \begin{tabular}{|cccccccc||cccccccc|}
    \hline
    EoS   & $f_{\mathrm{peak}}$ & $f_{2-0}$  & $f_{\mathrm{spiral}}$ & $m_1$ & $m_2$ & $M_{\rm chirp}$ &  $m_1/m_2$  & EoS  & $f_{\mathrm{peak}}$ & $f_{2-0}$  & $f_{\mathrm{spiral}}$ & $m_1$ & $m_2$ & $M_{\rm chirp}$ & $m_1/m_2$    \\
  & (kHz) & (kHz) & (kHz) & ($M_\odot$)   & ($M_\odot$)  &($M_\odot$) &  &   & (kHz) & (kHz) & (kHz) & ($M_\odot$)  &($M_\odot$)  &  ($M_\odot$) &     \\ \hline \hline
apr   & 3.46            & 2.33           & 2.68              & 1.35 & 1.35 & 1.18   & 1.00 & nl3  & 2.13            & -              & 1.55              & 1.30 & 1.40 & 1.17   & 0.93 \\\hline
bhblp & 2.95            & -              & 2.05              & 1.50 & 1.50 & 1.31   & 1.00 & nl3  & 2.37            & 1.39           & 1.79              & 1.60 & 1.60 & 1.39   & 1.00 \\\hline
bhblp & 2.65            & -              & -                 & 1.20 & 1.50 & 1.17   & 0.80 & nl3  & 2.04            & -              & 1.47              & 1.20 & 1.20 & 1.04   & 1.00 \\\hline
bhblp & 2.61            & 1.55           & 1.87              & 1.35 & 1.35 & 1.18   & 1.00 & nl3  & 2.07            & -              & -                 & 1.20 & 1.60 & 1.20   & 0.75 \\\hline
bhblp & 2.43            & -              & 1.81              & 1.20 & 1.20 & 1.04   & 1.00 & sfho & 3.07            & 1.85           & -                 & 1.20 & 1.35 & 1.11   & 0.89 \\\hline
dd2   & 2.67            & 1.54           & 1.94              & 1.35 & 1.50 & 1.24   & 0.90 & sfho & 3.30            & 2.15           & -                 & 1.20 & 1.60 & 1.20   & 0.75 \\\hline
dd2   & 3.08            & 2.18           & -                 & 1.65 & 1.65 & 1.44   & 1.00 & sfho & 3.31            & 2.34           & -                 & 1.30 & 1.50 & 1.22   & 0.87 \\\hline
dd2   & 2.62            & -              & -                 & 1.20 & 1.80 & 1.27   & 0.67 & sfho & 3.24            & 2.21           & -                 & 1.30 & 1.40 & 1.17   & 0.93 \\\hline
dd2   & 2.66            & 1.56           & 2.06              & 1.40 & 1.40 & 1.22   & 1.00 & sfho & 3.28            & 2.13           & 2.48              & 1.35 & 1.35 & 1.18   & 1.00 \\\hline
dd2   & 2.77            & 1.72           & 2.13              & 1.50 & 1.50 & 1.31   & 1.00 & sfho & 3.39            & -              & 2.51              & 1.40 & 1.40 & 1.22   & 1.00 \\\hline
dd2   & 2.80            & 1.72           & -                 & 1.35 & 1.80 & 1.35   & 0.75 & sfho & 2.99            & 1.70           & 2.21              & 1.20 & 1.20 & 1.04   & 1.00 \\\hline
dd2   & 2.50            & -              & -                 & 1.20 & 1.60 & 1.20   & 0.75 & sfho & 3.13            & 1.96           & -                 & 1.20 & 1.50 & 1.17   & 0.80 \\\hline
dd2   & 2.96            & 1.97           & 2.32              & 1.60 & 1.60 & 1.39   & 1.00 & sfhx & 3.41            & 2.31           & 2.56              & 1.50 & 1.50 & 1.31   & 1.00 \\\hline
dd2   & 2.55            & 1.41           & 1.86              & 1.30 & 1.40 & 1.17   & 0.93 & sfhx & 3.31            & 2.23           & 2.56              & 1.45 & 1.45 & 1.26   & 1.00 \\\hline
dd2   & 2.54            & 1.29           & 1.95              & 1.30 & 1.30 & 1.13   & 1.00 & sfhx & 2.91            & 1.81           & 2.06              & 1.20 & 1.35 & 1.11   & 0.89 \\\hline
dd2   & 2.61            & -              & -                 & 1.20 & 1.50 & 1.17   & 0.80 & sfhx & 2.81            & 1.55           & 2.01              & 1.20 & 1.20 & 1.04   & 1.00 \\\hline
dd2   & 2.94            & 1.98           & -                 & 1.50 & 1.80 & 1.43   & 0.83 & sfhx & 2.85            & 1.86           & -                 & 1.20 & 1.50 & 1.17   & 0.80 \\\hline
dd2   & 2.49            & 1.26           & 1.76              & 1.20 & 1.35 & 1.11   & 0.89 & sfhx & 3.06            & -              & -                 & 1.20 & 1.60 & 1.20   & 0.75 \\\hline
dd2   & 2.41            & 1.20           & 1.80              & 1.20 & 1.20 & 1.04   & 1.00 & sfhx & 3.20            & 2.11           & -                 & 1.35 & 1.50 & 1.24   & 0.90 \\\hline
dd2   & 2.60            & 1.47           & 1.98              & 1.35 & 1.35 & 1.18   & 1.00 & sfhx & 3.16            & 2.02           & 2.43              & 1.40 & 1.40 & 1.22   & 1.00 \\\hline
dd2   & 2.90            & 1.95           & -                 & 1.50 & 1.70 & 1.39   & 0.88 & sfhx & 3.08            & 1.86           & 2.34              & 1.35 & 1.35 & 1.18   & 1.00 \\\hline
ls220 & 3.06            & 2.07           & 2.54              & 1.40 & 1.40 & 1.22   & 1.00 & shen & 2.19            & -              & -                 & 1.20 & 1.50 & 1.17   & 0.80 \\\hline
ls220 & 2.59            & 1.42           & 1.77              & 1.20 & 1.20 & 1.04   & 1.00 & sly4 & 3.33            & 2.18           & 2.55              & 1.35 & 1.35 & 1.18   & 1.00 \\\hline
ls220 & 3.43            & 2.32           & 2.59              & 1.50 & 1.50 & 1.31   & 1.00 & tm1  & 2.23            & 1.16           & 1.43              & 1.20 & 1.50 & 1.17   & 0.80 \\\hline
ls220 & 2.85            & -              & -                 & 1.20 & 1.50 & 1.17   & 0.80 & tm1  & 2.90            & 1.61           & 2.11              & 1.70 & 1.70 & 1.48   & 1.00 \\\hline
ls220 & 2.87            & 1.72           & 2.18              & 1.35 & 1.35 & 1.18   & 1.00 & tm1  & 2.44            & 1.32           & 1.64              & 1.20 & 1.80 & 1.27   & 0.67 \\\hline
ls375 & 2.84            & 1.91           & -                 & 1.70 & 1.70 & 1.48   & 1.00 & tm1  & 2.61            & 1.67           & 2.01              & 1.60 & 1.60 & 1.39   & 1.00 \\\hline
ls375 & 2.39            & 1.33           & 1.76              & 1.35 & 1.35 & 1.18   & 1.00 & tm1  & 2.56            & 1.30           & -                 & 1.35 & 1.80 & 1.35   & 0.75 \\\hline
ls375 & 3.07            & 2.11           & 2.42              & 1.80 & 1.80 & 1.57   & 1.00 & tm1  & 2.73            & 1.83           & 2.10              & 1.65 & 1.65 & 1.44   & 1.00 \\\hline
ls375 & 2.41            & -              & -                 & 1.20 & 1.50 & 1.17   & 0.80 & tm1  & 2.25            & -              & 1.65              & 1.35 & 1.35 & 1.18   & 1.00 \\\hline
ls375 & 2.23            & -              & 1.65              & 1.20 & 1.20 & 1.04   & 1.00 & tm1  & 2.25            & 1.23           & 1.54              & 1.30 & 1.40 & 1.17   & 0.93 \\\hline
ls375 & 2.56            & 1.50           & 1.83              & 1.50 & 1.50 & 1.31   & 1.00 & tm1  & 2.12            & -              & 1.52              & 1.20 & 1.20 & 1.04   & 1.00 \\\hline
ls375 & 2.69            & 1.66           & 2.05              & 1.60 & 1.60 & 1.39   & 1.00 & tm1  & 2.52            & 1.49           & 1.79              & 1.50 & 1.50 & 1.31   & 1.00 \\\hline
nl3   & 2.58            & -              & 1.81              & 1.80 & 1.80 & 1.57   & 1.00 & tm1  & 2.34            & 1.28           & 1.57              & 1.35 & 1.50 & 1.24   & 0.90 \\\hline
nl3   & 2.34            & 1.38           & -                 & 1.35 & 2.00 & 1.42   & 0.68 & tm1  & 2.26            & 1.18           & -                 & 1.20 & 1.60 & 1.20   & 0.75 \\\hline
nl3   & 2.08            & -              & -                 & 1.20 & 1.50 & 1.17   & 0.80 & tm1  & 2.17            & -              & 1.46              & 1.20 & 1.35 & 1.11   & 0.89 \\\hline
nl3   & 2.20            & -              & 1.60              & 1.35 & 1.50 & 1.24   & 0.90 & tm1  & 2.74            & 1.71           & 1.87              & 1.50 & 1.80 & 1.43   & 0.83 \\\hline
nl3   & 2.19            & -              & -                 & 1.20 & 1.80 & 1.27   & 0.67 & tm1  & 2.33            & 1.20           & 1.73              & 1.40 & 1.40 & 1.22   & 1.00 \\\hline
nl3   & 2.15            & -              & 1.60              & 1.35 & 1.35 & 1.18   & 1.00 & tm1  & 2.20            & -              & 1.64              & 1.30 & 1.30 & 1.13   & 1.00 \\\hline
nl3   & 2.27            & 1.33           & -                 & 1.35 & 1.80 & 1.35   & 0.75 & tma  & 2.15            & 1.15           & 1.64              & 1.20 & 1.20 & 1.04   & 1.00 \\\hline
nl3   & 2.07            & -              & 1.30              & 1.20 & 1.35 & 1.11   & 0.89 & tma  & 2.97            & 1.63           & 2.18              & 1.60 & 1.60 & 1.39   & 1.00 \\\hline
nl3   & 2.69            & -              & 1.94              & 1.80 & 2.00 & 1.65   & 0.90 & tma  & 2.75            & 1.84           & 2.13              & 1.55 & 1.55 & 1.35   & 1.00 \\\hline
nl3   & 2.36            & 1.45           & -                 & 1.50 & 1.80 & 1.43   & 0.83 & tma  & 2.33            & 1.26           & -                 & 1.20 & 1.50 & 1.17   & 0.80 \\\hline
nl3   & 2.79            & 1.83           & 2.10              & 1.90 & 1.90 & 1.65   & 1.00 & tma  & 2.38            & 1.21           & 1.76              & 1.35 & 1.35 & 1.18   & 1.00 \\\hline
nl3   & 2.33            & -              & 1.60              & 1.50 & 1.50 & 1.31   & 1.00 & tma  & 2.73            & 1.82           & 2.21              & 1.50 & 1.50 & 1.31   & 1.00 \\ \hline
\end{tabular}
\label{table:cfc/sph-data}
\end{table*}

\begin{table}[h]
  \centering
\caption{Extracted frequency $f_{\rm peak}$,  along with individual masses $m_1$, $m_2$, chirp mass $M_{\rm chirp}$ and mass ratio $q$ for the models
comprising the  subset of the CoRe GW catalogue used in the present study
(in addition, $f_{\rm gw}$ is the GW frequency of the binary system at the
start of the simulation, as defined in the CoRe GW catalogue).}
\begin{tabular}{|c|c|c|c|c|c|c|}
\hline
EoS & \textbf{$f_{\mathrm{peak}}$}   & \textbf{$f_{\mathrm{gw}}$}  & \textbf{$m_1$}  & \textbf{$m_2$} & \textbf{$M_\mathrm{{chirp}}$}  &  $q$ \\ 
 & (kHz) &(Hz)
& ($M_\odot$)& ($M_\odot$)&
($M_\odot$) &  \\ \hline
ALF2 & 2.38  & 491.40 & 1.10 & 1.40 & 1.08   & 0.79 \\ \hline
ALF2 & 2.75  & 454.79 & 1.35 & 1.35 & 1.18   & 1.00 \\ \hline
ALF2 & 2.75  & 419.69 & 1.22 & 1.53 & 1.19   & 0.80 \\ \hline
ALF2 & 2.75  & 422.82 & 1.38 & 1.38 & 1.20   & 1.00 \\ \hline
ENG  & 3.00  & 454.57 & 1.35 & 1.35 & 1.18   & 1.00 \\ \hline
H4   & 2.54  & 410.94 & 1.00 & 1.75 & 1.14   & 0.57 \\ \hline
H4   & 2.56  & 410.86 & 1.10 & 1.65 & 1.17   & 0.67 \\ \hline
H4   & 2.41  & 454.55 & 1.35 & 1.35 & 1.18   & 1.00 \\ \hline
H4   & 2.47  & 409.87 & 1.22 & 1.53 & 1.19   & 0.80 \\ \hline
H4   & 2.53  & 408.95 & 1.38 & 1.38 & 1.20   & 1.00 \\ \hline
H4   & 2.86  & 424.99 & 1.50 & 1.50 & 1.31   & 1.00 \\ \hline
MPA1 & 2.81  & 454.55 & 1.35 & 1.35 & 1.18   & 1.00 \\ \hline
MS1b & 1.88  & 389.50 & 1.00 & 1.50 & 1.06   & 0.67 \\ \hline
MS1b & 2.08  & 490.42 & 1.10 & 1.40 & 1.08   & 0.79 \\ \hline
MS1b & 1.96  & 406.44 & 1.00 & 1.75 & 1.14   & 0.57 \\ \hline
MS1b & 2.15  & 402.75 & 0.94 & 1.94 & 1.16   & 0.49 \\ \hline
MS1b & 1.97  & 406.58 & 1.10 & 1.65 & 1.17   & 0.67 \\ \hline
MS1b & 2.09  & 418.48 & 1.35 & 1.35 & 1.18   & 1.00 \\ \hline
MS1b & 2.01  & 406.59 & 1.22 & 1.53 & 1.19   & 0.80 \\ \hline
MS1b & 2.09  & 407.98 & 1.38 & 1.38 & 1.20   & 1.00 \\ \hline
MS1b & 2.22  & 419.74 & 1.50 & 1.50 & 1.31   & 1.00 \\ \hline
SLy  & 2.94  & 407.76 & 1.00 & 1.50 & 1.06   & 0.67 \\ \hline
SLy  & 2.94  & 392.39 & 1.11 & 1.36 & 1.07   & 0.82 \\ \hline
SLy  & 2.79  & 491.39 & 1.10 & 1.40 & 1.08   & 0.79 \\ \hline
SLy  & 3.06  & 471.74 & 1.20 & 1.40 & 1.13   & 0.86 \\ \hline
SLy  & 3.40  & 453.15 & 1.35 & 1.35 & 1.18   & 1.00 \\ \hline
SLy  & 3.46  & 426.13 & 1.22 & 1.53 & 1.19   & 0.80 \\ \hline
\end{tabular}
\label{table:CORE-data}
\end{table}

\begin{table*}[h]
\centering
\caption{Regression information for the empirical relation of the form (\ref{fRM}).}
\begin{tabular}{|c|c|c|c|c|c|c|c|c|c|c|}
\hline
\textbf{$f_j$} & \textbf{$b_0$} & \textbf{$b_1$} & \textbf{$b_2$} & \textbf{$b_3$} &\textbf{$b_4$} &\textbf{$b_5$} & \textbf{$R^2$} & max res & mean res & $\sigma$ res \\ \hhline{|=|=|=|=|=|=|=|=|=|=|=|} 
\multicolumn{11}{| l |}{$f_{\mathrm{peak}}$ (equal masses)} \\ \hline
\textbf{$R_{1.2}$} &   18.203   &   -1.505   &  -1.944  &   -0.16   &   0.107   &  0.057   &  0.893   &  0.257  &  0.092  &  0.06  \\ \hline
\textbf{$R_{1.4}$} &   16.013   &   -1.092   &  -1.649  &   0.104   &   0.031   &  0.049   &  0.93   &  0.227  &  0.073  &  0.052  \\ \hline
\textbf{$R_{1.6}$} &   13.822   &   -0.576   &  -1.375  &   0.479   &   -0.073   &  0.044   &  0.956   &  0.196  &  0.056  &  0.044  \\ \hline
\textbf{$R_{1.8}$} &   12.168   &   -0.049   &  -1.205  &   0.954   &   -0.197   &  0.044   &  0.953   &  0.215  &  0.055  &  0.048  \\ 
\hhline{|=|=|=|=|=|=|=|=|=|=|=|} 
\multicolumn{11}{| l |}{$f_{\mathrm{peak}}$ (all masses)} \\ \hline
\textbf{$R_{1.2}$} &   16.91   &   -1.896   &  -1.708  &   0.2   &   0.068   &  0.05   &  0.88   &  0.374  &  0.096  &  0.07  \\ \hline
\textbf{$R_{1.4}$} &   14.819   &   -1.474   &  -1.43  &   0.414   &   -0.0   &  0.043   &  0.916   &  0.337  &  0.078  &  0.062  \\ \hline
\textbf{$R_{1.6}$} &   12.696   &   -0.935   &  -1.17  &   0.713   &   -0.092   &  0.037   &  0.943   &  0.298  &  0.062  &  0.053  \\ \hline
\textbf{$R_{1.8}$} &   10.942   &   -0.369   &  -0.987  &   1.095   &   -0.201   &  0.036   &  0.948   &  0.247  &  0.06  &  0.05  \\ 
\hhline{|=|=|=|=|=|=|=|=|=|=|=|} 
\multicolumn{11}{| l |}{$f_{\mathrm{2-0}}$ (equal masses)} \\ \hline
\textbf{$R_{1.2}$} &   12.607   &   3.074   &  -1.791  &   -0.928   &   -0.012   &  0.058   &  0.718   &  0.366  &  0.114  &  0.077  \\ \hline
\textbf{$R_{1.4}$} &   10.859   &   3.586   &  -1.571  &   -0.706   &   -0.087   &  0.053   &  0.79   &  0.306  &  0.096  &  0.069  \\ \hline
\textbf{$R_{1.6}$} &   8.943   &   4.059   &  -1.332  &   -0.358   &   -0.182   &  0.048   &  0.849   &  0.229  &  0.08  &  0.061  \\ \hline
\textbf{$R_{1.8}$} &   7.797   &   4.773   &  -1.256  &   0.285   &   -0.357   &  0.055   &  0.86   &  0.269  &  0.075  &  0.061  \\ 
\hhline{|=|=|=|=|=|=|=|=|=|=|=|} 
\multicolumn{11}{| l |}{$f_{\mathrm{2-0}}$ (all masses)} \\ \hline
\textbf{$R_{1.2}$} &   13.237   &   3.278   &  -1.894  &   -0.504   &   -0.107   &  0.066   &  0.785   &  0.383  &  0.105  &  0.081  \\ \hline
\textbf{$R_{1.4}$} &   11.549   &   3.76   &  -1.683  &   -0.26   &   -0.184   &  0.061   &  0.841   &  0.324  &  0.088  &  0.072  \\ \hline
\textbf{$R_{1.6}$} &   9.586   &   4.09   &  -1.427  &   0.048   &   -0.261   &  0.055   &  0.885   &  0.252  &  0.075  &  0.061  \\ \hline
\textbf{$R_{1.8}$} &   8.007   &   4.356   &  -1.241  &   0.558   &   -0.375   &  0.054   &  0.896   &  0.258  &  0.072  &  0.057  \\ \hline
\hhline{|=|=|=|=|=|=|=|=|=|=|=|} 
\multicolumn{11}{| l |}{$f_{\mathrm{spiral}}$ (equal masses)} \\ \hline
\textbf{$R_{1.2}$} &   10.565   &   1.013   &  -1.185  &   -0.184   &   -0.052   &  0.038   &  0.788   &  0.422  &  0.097  &  0.081  \\ \hline
\textbf{$R_{1.4}$} &   8.687   &   1.398   &  -0.934  &   0.096   &   -0.13   &  0.032   &  0.835   &  0.37  &  0.083  &  0.075  \\ \hline
\textbf{$R_{1.6}$} &   7.019   &   1.756   &  -0.721  &   0.468   &   -0.222   &  0.028   &  0.872   &  0.306  &  0.074  &  0.065  \\ \hline
\textbf{$R_{1.8}$} &   6.264   &   1.929   &  -0.645  &   0.881   &   -0.311   &  0.03   &  0.877   &  0.286  &  0.075  &  0.061  \\ \hline
\hhline{|=|=|=|=|=|=|=|=|=|=|=|} 
\multicolumn{11}{| l |}{$f_{\mathrm{spiral}}$ (all masses)} \\ \hline
\textbf{$R_{1.2}$} &   8.942   &   0.926   &  -0.926  &   -0.069   &   -0.061   &  0.028   &  0.773   &  0.438  &  0.109  &  0.079  \\ \hline
\textbf{$R_{1.4}$} &   7.356   &   1.321   &  -0.719  &   0.218   &   -0.141   &  0.024   &  0.814   &  0.383  &  0.1  &  0.07  \\ \hline
\textbf{$R_{1.6}$} &   6.107   &   1.666   &  -0.567  &   0.596   &   -0.234   &  0.022   &  0.845   &  0.316  &  0.092  &  0.063  \\ \hline
\textbf{$R_{1.8}$} &   5.846   &   1.75   &  -0.555  &   1.002   &   -0.316   &  0.026   &  0.846   &  0.27  &  0.089  &  0.066  \\ \hline
\end{tabular}
\label{table:fRM}
\end{table*}

\begin{table*}[h]
\centering
\caption{Regression information for the empirical relation of the form (\ref{RfM}).}
\begin{tabular}{|c|c|c|c|c|c|c|c|c|c|c|}
\hline
\textbf{$R_{\mathrm{x}}$} & \textbf{$b_0$} & \textbf{$b_1$} & \textbf{$b_2$} & \textbf{$b_3$} &\textbf{$b_4$} &\textbf{$b_5$} & \textbf{$R^2$} & max res & mean res & $\sigma$ res \\ \hhline{|=|=|=|=|=|=|=|=|=|=|=|} 
\multicolumn{11}{| l |}{$R_{1.2}^{<}$ (equal masses)} \\ \hline
\textbf{$f_{\mathrm{peak}}$} &   52.201   &   -29.769   &  -15.398  &   8.918   &   3.333   &  1.832   &  0.945   &  0.52  &  0.191  &  0.131  \\ \hline
\textbf{$f_{\mathrm{2-0}}$} &   29.638   &   -19.343   &  -7.525  &   9.087   &   0.521   &  1.454   &  0.871   &  0.595  &  0.259  &  0.157  \\ \hline
\textbf{$f_{\mathrm{spiral}}$} &   41.603   &   -14.538   &  -19.426  &   0.955   &   6.832   &  2.257   &  0.905   &  0.803  &  0.245  &  0.181  \\ \hline
\multicolumn{11}{| l |}{$R_{1.2}^{<}$ (all masses)} \\ \hline
\textbf{$f_{\mathrm{peak}}$ } &   56.906   &   -37.252   &  -15.701  &   11.756   &   3.638   &  1.83   &  0.951   &  0.526  &  0.19  &  0.134  \\ \hline
\textbf{$f_{\mathrm{2-0}}$} &   31.374   &   -19.386   &  -9.852  &   9.845   &   -0.763   &  2.753   &  0.918   &  0.646  &  0.218  &  0.167  \\ \hline
\textbf{$f_{\mathrm{spiral}}$} &   38.805   &   -20.695   &  -12.612  &   5.561   &   4.366   &  1.212   &  0.898   &  0.737  &  0.275  &  0.18  \\ \hline
\multicolumn{11}{| l |}{$R_{1.4}^{<}$ (equal masses)} \\ \hline
\textbf{$f_{\mathrm{peak}}$} &    51.229   &   -30.463   &  -14.143  &   9.46   &   3.09   &  1.612   &  0.966   &  0.412  &  0.147  &  0.108 \\ \hline
\textbf{$f_{\mathrm{2-0}}$} &    28.249   &   -17.137   &  -7.399  &   7.988   &   0.9   &  1.22   &  0.912   &  0.506  &  0.205  &  0.147 \\ \hline
\textbf{$f_{\mathrm{spiral}}$} &   40.407   &   -14.521   &  -17.99  &   1.256   &   6.435   &  1.958   &  0.926   &  0.731  &  0.212  &  0.167   \\ \hline
\multicolumn{11}{| l |}{$R_{1.4}^{<}$ (all masses)} \\ \hline
\textbf{$f_{\mathrm{peak}}$ } &    55.809   &   -37.642   &  -14.473  &   12.15   &   3.41   &  1.609   &  0.968   &  0.493  &  0.154  &  0.109  \\ \hline
\textbf{$f_{\mathrm{2-0}}$} &    30.105   &   -17.84   &  -9.364  &   8.874   &   -0.137   &  2.305   &  0.942   &  0.564  &  0.18  &  0.144  \\ \hline
\textbf{$f_{\mathrm{spiral}}$} &   37.684   &   -20.594   &  -11.299  &   5.687   &   4.143   &  0.89   &  0.914   &  0.676  &  0.25  &  0.169 \\ \hline
\multicolumn{11}{| l |}{$R_{1.6}$ (equal masses)} \\ \hline
\textbf{$f_{\mathrm{peak}}$} &   41.316   &   -16.654   &  -12.458  &   3.722   &   2.936   &  1.269   &  0.969   &  0.462  &  0.139  &  0.108   \\ \hline
\textbf{$f_{\mathrm{2-0}}$} &   15.271   &   4.123   &  -6.661  &   -1.188   &   1.23   &  0.783   &  0.942   &  0.465  &  0.186  &  0.13 \\ \hline
\textbf{$f_{\mathrm{spiral}}$} &  40.081   &   -18.359   &  -15.205  &   3.98   &   5.187   &  1.544   &  0.941   &  0.706  &  0.197  &  0.152 \\ \hline
\multicolumn{11}{| l |}{$R_{1.6}$ (all masses)} \\ \hline
\textbf{$f_{\mathrm{peak}}$ } &   43.796   &   -19.984   &  -12.921  &   4.674   &   3.371   &  1.26   &  0.969   &  0.526  &  0.144  &  0.117 \\ \hline
\textbf{$f_{\mathrm{2-0}}$} &   17.764   &   2.497   &  -8.797  &   -0.639   &   1.393   &  1.452   &  0.955   &  0.518  &  0.174  &  0.13   \\ \hline
\textbf{$f_{\mathrm{spiral}}$} &   30.762   &   -12.647   &  -8.704  &   3.081   &   3.225   &  0.414   &  0.926   &  0.674  &  0.236  &  0.147   \\ \hline
\multicolumn{11}{| l |}{$R_{1.8}^{>}$ (equal masses)} \\ \hline
\textbf{$f_{\mathrm{peak}}$} &   33.802   &   -3.069   &  -15.522  &   -1.439   &   4.112   &  1.605   &  0.951   &  0.276  &  0.107  &  0.067  \\ \hline
\textbf{$f_{\mathrm{2-0}}$} &    34.725   &   -15.096   &  -15.795  &   4.743   &   2.745   &  3.623   &  0.779   &  0.597  &  0.176  &  0.18 \\ \hline
\textbf{$f_{\mathrm{spiral}}$} &   55.934   &   -37.162   &  -17.139  &   7.961   &   9.897   &  -0.382   &  0.951   &  0.212  &  0.117  &  0.05   \\ \hline
\multicolumn{11}{| l |}{$R_{1.8}^{>}$ (all masses)} \\ \hline
\textbf{$f_{\mathrm{peak}}$ } &    28.796   &   -7.668   &  -6.631  &   0.516   &   3.478   &  -0.492   &  0.958   &  0.275  &  0.11  &  0.067 \\ \hline
\textbf{$f_{\mathrm{2-0}}$} &  0.747   &   24.015   &  -7.446  &   -6.192   &   -2.413   &  3.136   &  0.83   &  0.569  &  0.184  &  0.17 \\ \hline
\textbf{$f_{\mathrm{spiral}}$} &   54.468   &   -38.851   &  -13.993  &   9.305   &   8.453   &  -0.614   &  0.921   &  0.34  &  0.138  &  0.082   \\ \hline
\end{tabular}
\label{table:RfM}
\end{table*}

\end{document}